\documentclass[useAMS,usenatbib]{mn2e}
\usepackage{psfig}
\usepackage{amssymb}
\usepackage{color}
\usepackage{enumerate}
\usepackage{rotating}
\usepackage{amsmath}
\usepackage{mathrsfs}
\usepackage{ragged2e}
\usepackage{hyperref}
\usepackage{comment}

\newcommand{\be}{\begin{equation}}
\newcommand{\ee}{\end{equation}}
\newcommand{\bea}{\begin{eqnarray}}
\newcommand{\eea}{\end{eqnarray}}
\newcommand{\no}{\nonumber}

\title[The manifestation of secondary bias on galaxies]
{The manifestation of secondary halo bias on the galaxy population from IllustrisTNG300}

\author[Montero-Dorta et al.]{
\parbox[t]{\textwidth}{
Antonio D. Montero-Dorta$^{1}$\thanks{E-mail: amonterodorta@gmail.com}, M. Celeste Artale$^{2}$, L. Raul Abramo$^{1}$, Beatriz Tucci$^{1}$, Nelson Padilla$^{3,4}$, Gabriela Sato-Polito$^{5}$, Iv\'an Lacerna$^{6,7}$, Facundo Rodriguez$^{8,9}$, Raul E. Angulo$^{10,11}$} 
\vspace*{6pt} \\ 
$^1$ Departamento de F\'isica Matem\'atica, Instituto de F\'isica, Universidade de S\~ao Paulo, Rua do Mat\~ao 1371, CEP 05508-090, \\
S\~ao Paulo, Brazil \\
$^2$ Institut f\"ur Astro- und Teilchenphysik, Universit\"at Innsbruck, Technikerstrasse 25/8, 6020 Innsbruck, Austria \\
$^3$ Instituto Astrof\'isica, Pontificia Universidad Cat\'olica de Chile, Santiago, Chile \\
$^4$ Centro de Astro-Ingenier\'ia, Pontificia Universidad Cat\'olica de Chile, Santiago, Chile \\
$^5$ Department of Physics \& Astronomy, Johns Hopkins University,3400 N. Charles St., Baltimore, MD 21218, USA \\
$^6$ Instituto de Astronom\'ia y Ciencias Planetarias, Universidad de Atacama, Copayapu 485, Copiap\'o, Chile \\
$^7$ Instituto Milenio de Astrof\'isica, Av. Vicu\~na Mackenna 4860, Macul, Santiago, Chile \\
$^8$ Universidad Nacional de C\'ordoba. Observatorio Astron\'omico de C\'ordoba. C\'ordoba, Argentina \\
$^9$ CONICET. Instituto de Astronom\'ia Te\'orica y Experimental. Laprida 854, X5000BGR, C\'ordoba, Argentina \\
$^{10}$ Donostia International Physics Center (DIPC), Paseo Manuel de Lardiz\'abal, 4, 20018 Donostia-San Sebasti\'an, Spain \\
$^{11}$ IKERBASQUE, Basque Foundation for Science, 48013, Bilbao, Spain
\vspace{-0.4cm} 
}

\date{Accepted ---. Received ---;in original form --- \vspace{-0.3cm}}

\def\simlt{\lower.5ex\hbox{$\; \buildrel < \over \sim \;$}}
\def\simgt{\lower.5ex\hbox{$\; \buildrel > \over \sim \;$}}
\usepackage{graphicx}
\usepackage{rotating}

\definecolor{red}{rgb}{1,0,0}

\begin{document}

\bibliographystyle{mnras}

\maketitle

\begin{abstract}
We use the improved IllustrisTNG300 magneto-hydrodynamical cosmological simulation to revisit the effect that secondary halo bias has on the clustering of the central galaxy population. With a side length of 205 $h^{-1}$Mpc and significant improvements on the sub-grid model with respect to the previous Illustris boxes, IllustrisTNG300 allows us to explore the dependencies of galaxy clustering over a large cosmological volume and wide halo-mass range. We show, at high statistical significance, that the halo assembly bias signal (i.e., the secondary dependence of halo bias on halo formation redshift) manifests itself on the clustering of the central galaxy population when this is split by stellar mass, colour, specific star formation rate, and surface density. A significant detection is also obtained for galaxy size: at fixed halo mass, larger central galaxies are more tightly clustered than smaller central galaxies in haloes of mass M$_{\rm vir} \lesssim 10^{12.5}$ $h^{-1}$M$_{\odot}$. This effect, however, seems to be uncorrelated with halo formation time, unlike the rest of the secondary dependencies analysed. We also explore the transmission of the halo spin bias signal, i.e., the secondary dependence of halo bias on halo spin. Although galaxy spin retains little information about the total spin of the halo, the correlation is enough to produce a significant galaxy spin bias signal. We discuss possible ways to probe the spin bias effects with observations.

\end{abstract}

\begin{keywords}

methods: numerical - galaxies: formation - galaxies: haloes - dark matter - large-scale structure of Universe - cosmology: theory

\end{keywords}

\section{Introduction}
\label{sec:intro}

In the standard model of cosmology, dark matter (DM) clusters along density peaks that were generated during inflation, collapsing later on to form {\it{dark-matter haloes}}. It is inside these collapsing structures that galaxies form, when gas falls into their potential wells (e.g., \citealt{white1991}). The relationship between galaxies, haloes, and the underlying matter distribution is therefore crucial to our understanding of the galaxy formation process, and to our ability to test cosmological models against observations. 

The clustering of dark-matter haloes is commonly characterised by {\it{halo bias}}, which can be broadly defined as the relation between the spatial distribution of haloes and the underlying matter density field. In its simplest description, the linear halo bias can be assumed to depend only on halo mass, with more massive haloes being more strongly clustered than less massive haloes (e.g., \citealt{Kaiser1984,ShethTormen1999}). However, halo clustering is a very complex process that is known to depend on a variety of secondary halo properties. Among these secondary dependencies, the one that has drawn more attention is the dependence on the assembly history of haloes, an effect called {\it{halo assembly bias}}{\footnote{Throughout this paper, halo assembly bias is considered a particular case of secondary halo bias where the secondary halo property considered is directly related to halo formation time or halo accretion history.}}. Lower-mass haloes that assemble a significant portion of their mass early on are more tightly clustered than haloes that assemble at later times, {\it{at fixed halo mass}} (see, e.g., \citealt{Sheth2004,gao2005, wechsler2006,Gao2007,Angulo2008,li2008,Lazeyras2017,salcedo2018,han2018,Mao2018, SatoPolito2019, Johnson2019}). At higher halo masses (M$_{\rm vir} \gtrsim 10^{14}$ $h^{-1}$M$_{\odot}$), the signal seems to depend strongly on the definition of halo age \citep{Chue2018, li2008}. Besides halo assembly bias, a number of other secondary dependencies have been identified for halo clustering (on, e.g., concentration, spin, shape, environment), but a comprehensive theory for the physical origins of these effects is yet to be established (see, e.g., \citealt{dalal2008,Paranjape2018,Ramakrishnan2019}).

A question that has stirred up debate in recent years is whether the above secondary halo bias effects manifest themselves on the galaxy population, and, if so, whether they can actually be detected with current data. The term {\it{galaxy assembly bias}} is often used to refer to the dependence of galaxy clustering on secondary halo properties beyond halo mass (see, e.g., \citealt{Croton2006,Zhu2006,Zentner2014,Lacerna2014, Hearin2015, Zentner2016,Miyatake2016,Zu2016, Lin2016, Romano2017,MonteroDorta2017B,Niemiec2018,Walsh2019}). This signal could be simply a direct manifestation of secondary halo bias on the galaxy population, but could also include other contributions that are not directly related to halo formation. In the context of halo occupation distribution (HOD) modelling, galaxy assembly bias is described as the combination of halo assembly bias and the so-called {\it{occupancy variations}}, i.e., the dependencies of the galaxy content of haloes on secondary halo properties at fixed halo mass (see, e.g., \citealt{Artale2018,Zehavi2018,Bose2019}).

With the Sloan Digital Sky Server (SDSS, \citealt{York2000}) at low redshift, \cite{Miyatake2016} and \cite{More2016} claimed the detection of the so-called concentration bias on galaxy clusters, i.e., the secondary dependence of clustering on concentration (which is related to halo formation time\footnote{Note that, nevertheless, the secondary halo bias signal found for halo age is different to that found for concentration (see, e.g., \citealt{salcedo2018,Chue2018,SatoPolito2019}).}), for high-mass haloes. This claim was subsequently refuted by the evidence of projection effects that affected significantly the identification of cluster members (\citealt{Zu2016,Sunayama2019}). Also using the SDSS, \cite{Lin2016} found little evidence of any dependence of galaxy clustering on star formation history (SFH) beyond what is expected from the measured halo-mass difference. These results appear in some tension with findings for luminous red galaxies (LRGs) at $z=0.55$ (see \citealt{MonteroDorta2017B} and \citealt{Niemiec2018}, who used the Baryon Oscillation Spectroscopic Survey, BOSS, \citealt{Dawson2013}). In these works, a dependence of the amplitude of clustering on SFH is shown for LRGs, which also seems to occur at fixed halo mass (within the weak-lensing errors). Several other works have addressed the galaxy assembly bias question, producing mixed results. \cite{Lacerna2014}, for instance, used both SDSS and mock galaxies to claim a weak, although significant detection of assembly bias for central galaxies. 

Despite these contradictory results, a convincing proof of the existence of galaxy assembly bias seems attainable with upcoming cosmological surveys, which will reduce significantly the errors on the weak-lensing and clustering measurements, and will allow a better identification and characterisation of galaxy clusters. Note that, although most of the previous galaxy assembly bias analyses have focused on properties related to halo accretion history, it is conceivable that other secondary halo bias dependencies leave an imprint on the galaxy population. The dependence on halo spin seems particularly appealing
in this context, given its large effect at the high-mass end (see, e.g., \citealt{SatoPolito2019, Johnson2019}).

\begin{figure}
\begin{center}
\includegraphics[width=\columnwidth]{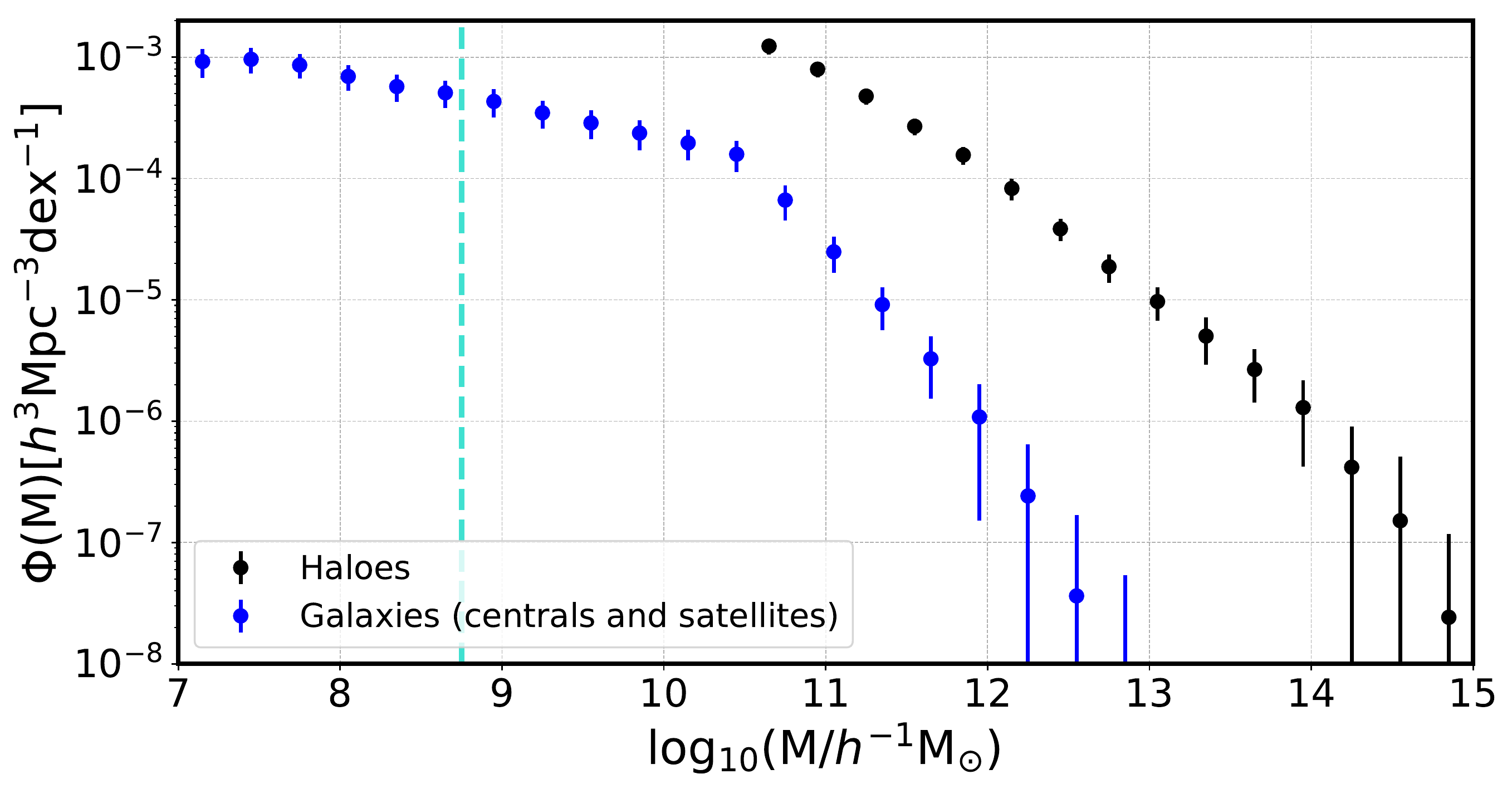}\hfill
\caption{The halo and stellar mass functions in the IllustrisTNG300 simulation at $z=0$. In this paper, only haloes of $\log_{10} ({\rm M_{vir}}/h^{-1} {\rm M_{\odot}}) > 10.5$ are considered. In this plot, the mass labelled ``M" corresponds to the virial halo mass (M$_{\rm vir}$) for haloes and the stellar mass (M$_{*}$) for galaxies. Error bars show the box-to-box variation for a set of 64 sub-boxes. Note that the mass resolution is 
$4.0 \times 10^7 ~h^{-1} {\rm M_{\odot}}$ for dark-matter particles and $7.6 \times 10^6~h^{-1} {\rm M_{\odot}}$ for baryonic particles. The vertical dashed line marks our resolution limit of 50 gas particles for galaxies.}
\label{fig:mf}
\end{center}
\end{figure}

\begin{figure}
\begin{center}
\includegraphics[width=\columnwidth]{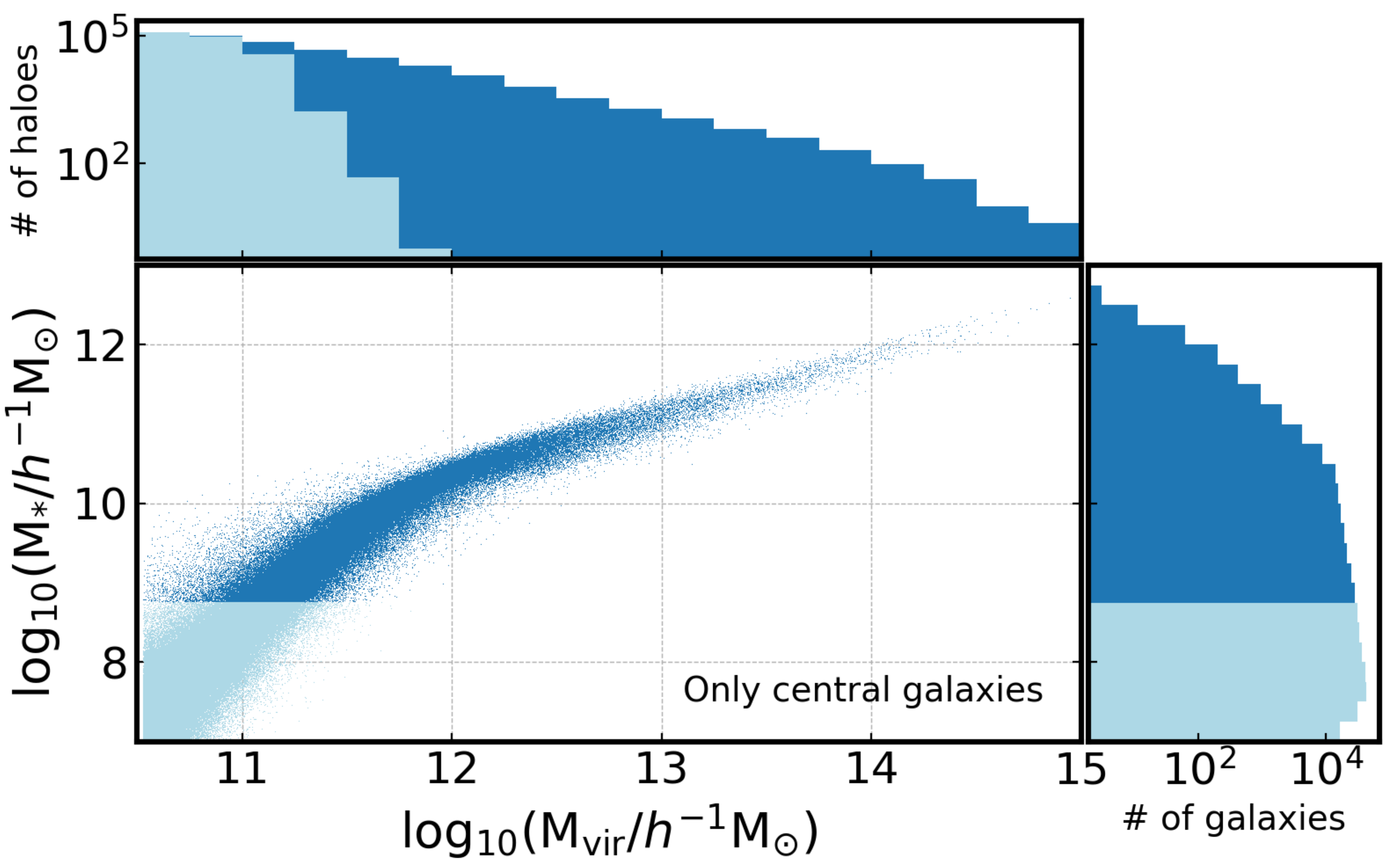}\hfill
\caption{The relation between stellar mass and halo mass for central galaxies in the IllustrisTNG300 simulation. The histograms above and to the right of the scatter plot show the distributions of halo virial mass and stellar mass, respectively, in logarithmic units. Lighter colours mark the distribution of galaxies (and their hosting haloes) below our resolution limit of 50 gas particles.}
\label{fig:smhmr}
\end{center}
\end{figure}

In this paper, we use the IllustrisTNG simulation at redshift $z=0$ to study how the secondary halo bias signal transmits to the galaxy population. IllustrisTNG is an ongoing suite of magneto-hydrodynamical cosmological simulations that model the formation and evolution of galaxies within the $\Lambda$CDM paradigm. Multiple refinements in the sub-grid model have improved significantly the performance of the simulation with respect to Illustris, in terms of reproducing important observational constraints such as the stellar mass function, along with halo--galaxy relations such as the stellar-to-halo mass relation (SHMR). These updates range from a new implementation of galactic winds, to black-hole-driven kinetic feedback at low accretion rates and the inclusion of magneto-hydrodynamics \citep{Weinberger2017,Pillepich2018}. In the context of assembly bias, hydro-dynamical simulations provide predictions that are less model-dependent than those coming from semi-analytic models (SAMs), even though these have proven to be excellent tools to investigate multiple aspects of the effect \citep[see e.g.,][]{Croton2006,Lacerna2018,Zehavi2018,Zehavi2019,Padilla2019,Contreras2019}. 

Among the different IllustrisTNG boxes, we choose the IllustrisTNG300 box, which presents an additional advantage as compared to previous Illustris versions. With a side length of 205 $h^{-1}$Mpc, it starts to be statistically comparable to some of the N-body numerical simulations that have been used in the context of the measurement of secondary halo bias, such as the MultiDark boxes \citep{Klypin2017}. This allows us to increase the statistical significance of the measurement presented in \citet{Xu2018}, who used the 75-$h^{-1}$Mpc Illustris-2 box. The volume and resolution of IllustrisTNG300 allow us to analyse the 
clustering of central galaxies in haloes within the mass range $10.5 \lesssim \log_{10} ({\rm M_{vir}}/ h^{-1} {\rm M_{\odot}}) \lesssim 14.5$.

The aforementioned analysis from \citet{Xu2018} shows that the halo assembly bias signal is indeed reflected on the clustering of the galaxy population (a prediction that, again, has not been convincingly confirmed with observations). Their results from Illustris-2 also expose a strong correlation between the stellar mass of central galaxies and the peak maximum circular velocity of their hosting haloes (V$_{\rm peak}$). The maximum circular velocity has, in fact, been proposed as a more efficient property in terms of encapsulating the assembly bias effect, as compared to halo mass (see, e.g., \citealt{Xu2018,zehavi2005,Zehavi2019}).

Using the 67-$h^{-1}$Mpc {\sc eagle} simulation, \cite{ChavesMontero2016} quantified the assembly bias clustering effect in 20$\%$ with respect to the standard mass-based sub-halo (hereafter {\it{subhalo}}) abundance matching (SHAM) framework. Several other aspects that are tightly related to assembly bias have been addressed with hydro-dynamical simulations. These include the effect of {\it{galactic conformity}}, a term that generally refers to the correlations in the colours and star formation rates of neighbouring galaxies (see \citealt{Bray2016} for an analysis on Illustris), and the aforementioned occupancy variations \citep{Artale2018,Bose2019}. The analysis on occupancy variations presented in \cite{Bose2019} can, in fact, be seen as complementary to our work. Their results, also obtained with IllustrisTNG300, indicate a dependence of the number of satellites on halo properties, at fixed halo mass. It is shown that haloes tend to harbour more satellites when they are less concentrated or younger, live in dense environments, and have higher angular momenta. The probability of hosting a central galaxy, on the other hand, is enhanced for low-mass high-concentration haloes and for low-mass haloes that live in overdense regions. 

The paper is organised as follows. Section~\ref{sec:sims} and \ref{sec:properties} provide, respectively, brief descriptions of both the simulation boxes and the halo and galaxy properties analysed in this work. 
The methodology for the computation of the relative bias between subsets of haloes/galaxies is explained in Section~\ref{sec:methodology}. The main results of our analysis in terms of the manifestation of secondary halo bias on the galaxy population are presented in Section~\ref{sec:results}. Finally, Section~\ref{sec:discussion} is devoted to discussing the implications of these results and providing a brief summary of the paper. The IllustrisTNG300 simulation adopts the standard $\Lambda$CDM cosmology  \citep{Planck2016}, with parameters $\Omega_{\rm m} = 0.3089$,  $\Omega_{\rm b} = 0.0486$, $\Omega_\Lambda = 0.6911$, $H_0 = 100\,h\, {\rm km\, s^{-1}Mpc^{-1}}$ with $h=0.6774$, $\sigma_8 = 0.8159$ and $n_s = 0.9667$.


\section{Simulations}
\label{sec:sims}

In this paper, we use the galaxy and dark-matter halo catalogues from 
{\it The Next Generation} Illustris  (IllustrisTNG)\footnote{\url{http://www.tng-project.org}}
magneto-hydrodynamical cosmological simulations, which represent an updated version of the Illustris simulations \citep{Vogelsberger2014a, Vogelsberger2014b, Genel2014}.
The IllustrisTNG simulations are performed with the {\sc arepo} moving-mesh code \citep{Springel2010} and include sub-grid models that account for radiative metal-line gas cooling, star formation, chemical enrichment from SNII, SNIa and AGB stars, stellar feedback, supermassive-black-hole formation with multi-mode quasar, and kinetic black-hole feedback. The main updates with respect to the Illustris simulation are: a new implementation of black-hole kinetic feedback at low accretion rates, a revised scheme for galactic winds, and the inclusion of magneto-hydrodynamics \citep[see][for further details]{Pillepich2018,Weinberger2017}.

In this work, we analyse the IllustrisTNG300-1 run and its {\it{dark-matter-only}} counterpart IllustrisTNG300-1-DMO (hereafter IllustrisTNG300 and IllustrisTNG300-DMO, respectively), which are the largest simulated boxes from the IllustrisTNG suite featuring the highest resolution level. These runs adopt a cubic box of side $205\,h^{-1}$~Mpc with periodic boundary conditions. The IllustrisTNG300 run follows the evolution of 2500$^3$ dark-matter particles of mass $4.0 \times 10^7 h^{-1} {\rm M_{\odot}}$, and 2500$^3$ gas cells of mass $7.6 \times 10^6 h^{-1} {\rm M_{\odot}}$. The IllustrisTNG300-DMO boxes contain 2500$^3$ dark-matter particles with mass $7.0\times10^7 h^{-1} {\rm M_{\odot}}$. 

The IllustrisTNG300 simulation has proven to be a powerful and self-consistent tool to investigate the distribution of galaxies and dark-matter haloes on large scales, one of the main focuses of our work. \cite{Springel2018} analysed the two-point correlation function of galaxies (and dark-matter haloes)
in IllustrisTNG300, finding good agreement with observations in terms of the stellar-mass and colour dependence of galaxy clustering. Also complementary to our analysis is the study presented in \cite{Bose2019}, who explored occupancy variations of dark-matter haloes at a fixed halo mass as a function of environment and secondary properties such as concentration, formation time, and angular momentum. Several other constraints related to the properties of galaxies have been shown to agree with observations, including the size distributions of star forming and quiescent galaxies \citep{Genel2018}, the evolution of the galaxy stellar mass function \citep{Pillepich2018}, and the location and shape of the red sequence and blue cloud of the $z=0$ SDSS galaxy population \citep{Nelson2018}. These results lay a solid foundation for our analysis.

\section{Halo and galaxy properties}
\label{sec:properties}

In IllustrisTNG300, dark-matter haloes are identified using a friends-of-friends (FOF) algorithm with a linking length of 0.2 times the mean inter-particle separation \citep{Davis1985}. The gravitationally bound substructures called subhaloes are subsequently identified using the SUBFIND algorithm \citep{Springel2001,Dolag2009}. 
Among all subhaloes, those containing a non-zero stellar component are considered galaxies. Each dark-matter halo typically contains multiple galaxies, including a central galaxy and several satellites, where the positions of centrals coincide with the FOF centres. 

In order to compute the formation time of haloes, we use the
subhalo merger trees that were obtained with the SUBLINK algorithm \citep{Rodriguez-Gomez2015} and are publicly available at the IllustrisTNG database.

In our analysis, we focus on the following halo properties:

\begin{itemize}
  \item Virial mass, M$_{\rm vir}$ [$h^{-1} {\rm M_{\odot}}$], computed by adding up the mass of all gas cells and particles enclosed within a sphere of radius R$_{\rm vir}$ that is defined so that the enclosed density equals 200 times the critical density. 

  \item Formation redshift, z$_\text{1/2}$, defined as the redshift at which half of the present-day halo mass has been accreted into a single subhalo for the first time. For this, we use the progenitors of the main branch of the subhalo merger tree computed with SUBLINK, which is initialized at z $ = 6$.
  
  \item Spin, $\lambda_{\rm halo}$, defined as in \cite{Bullock2001}:

  \begin{equation}\label{eq:spin}
    \lambda_{\rm halo} = \frac{\rm |J|}{\sqrt{2} {\rm M_{vir}} {\rm V_{vir}} {\rm R_{vir}}},
  \end{equation}
  where J is the angular momentum of the halo and V$_{\rm vir}$ is its circular velocity at the virial radius R$_{\rm vir}$. Note that some works use the \cite{Peebles1969} halo spin definition in the context of secondary bias (see, e.g., \citealt{Lacerna2012}), but this choice is known to have little impact on the secondary bias signal (\citealt{Johnson2019}). 

  \item Concentration, c$_{200}$, defined as
  
  \begin{equation}
  {\rm c_{200} = \frac{R_{vir}}{R_{s}}},
  \end{equation}
	where R$_s$ is the scale radius, obtained from fitting the dark-matter density profiles of individual haloes with an NFW profile \citep*{nfw1997}.
\end{itemize}

For the simulated galaxies (i.e., subhaloes with non-zero stellar components), we investigate the secondary bias effect coming from:

\begin{itemize}
    \item Star formation rate, SFR [${\rm M_{\odot}~yr^{-1}}$], computed as the sum of the star formation rate of all gas cells in each subhalo.
    \item Stellar mass, ${\rm M_\ast{}}$ [$h^{-1} {\rm M_{\odot}}$], defined as the total mass of all stellar particles bound to each subhalo.
    \item Specific star formation rate, sSFR [$h\,{\rm yr^{-1}}$], defined as sSFR = SFR/M$_\ast{}$.
    \item Stellar half-mass radius, R$_{1/2}^{(*)}$ [$h^{-1} \, {\rm kpc}$], defined as the comoving radius containing half of the stellar mass of each subhalo.
    \item Subhalo half-mass radius, R$_{1/2}^{(sh)}$ [$h^{-1} \, {\rm kpc}$], defined as the comoving radius containing half of the total (stellar+gas+DM) mass of each subhalo. 
    \item Galaxy colour (g-i), computed using the magnitudes provided at the IllustrisTNG database. The magnitudes are computed by summing up the luminosities of the stellar particles of each subhalo (based on the procedure of \citealt{Buser1978}).
    \item Galaxy total spin, $\lambda_{\rm galaxy}$, defined as in eq.~\ref{eq:spin} but using all the particles (DM + stellar + gas components) inside the stellar half-mass radius of the galaxy, R$_{1/2}^{(*)}$.
    \item Galaxy stellar spin, $\lambda_{*}$, defined as in eq.~\ref{eq:spin}, but using only the stellar component inside R$_{1/2}^{(*)}$.

    \item Surface density [$h~ {\rm M_{\odot}} \, {\rm kpc}^{-2}$], defined as the stellar mass divided by the square of the stellar half-mass radius.
    \item Velocity dispersion, $\sigma$ [${\rm km} \, {\rm s}^{-1}$], defined as the one-dimensional velocity dispersion of all the particles and cells of each subhalo. 
\end{itemize}

Figure~\ref{fig:mf} shows the halo and stellar mass functions in the IllustrisTNG300 $z=0$ box (including both centrals and satellites). In order to ensure good resolution, only haloes above $\log_{10} ({\rm M_{vir}}/h^{-1} {\rm M_{\odot}}) > 10.5$ and central galaxies with $\log_{10} ({\rm M_{*}}/h^{-1} {\rm M_{\odot}}) > 8.75$ are considered in this analysis. This imposes a threshold of at least $\sim$ 1000 particles per halo and $\sim$ 50 gas cells for central galaxies.

In Figure~\ref{fig:smhmr}, we show the relation between halo mass and the stellar mass of central galaxies in the simulation (the stellar-to-halo mass relation, SHMR), along with the distributions of these two quantities (in logarithmic scale).

\section{Methodology: Relative Bias Measurement}
\label{sec:methodology}

In order to quantify the dependence of clustering on a secondary property S, we measure the relative bias, $b_{\rm r}$, between a subset of objects selected according to S and all objects in the same primary bias property (B) bin. Note that in this analysis, the primary property B is always the halo virial mass, M$_{\rm vir}$, whereas S can be either a halo or a central galaxy property. We use subsets of 50$\%$ of the entire sample in each halo mass bin, following \cite{Xu2018}. This choice slightly attenuates the magnitude of the secondary bias effect with respect to the standard way of presenting the measurement (based on 25$\%$ quartiles), but it allows us to improve the statistics at the high-mass end significantly.  

In order to further increase the signal-to-noise in the computation of the relative bias, we use the cross-correlation with the entire sample (i.e., all mass bins), for both the S-subset and the B-bin. Namely, for a given halo-mass bin B: 

\begin{equation}
   b_{\rm r}({\rm r,S|B}) = \frac{\xi_{[{\rm S},all]}({\rm r, S})}{\xi_{[{\rm B},all]}({\rm r, B})},
   \label{eq:bias}
\end{equation}

\noindent where $\xi_{[{\rm S},all]}$ is the cross-correlation between all objects in the S-subset and all objects in the sample, and $\xi_{[{\rm B},all]}$ is the cross-correlation between all objects in the halo-mass bin and the entire sample. Note, again, that B$\rightarrow$M$_{\rm vir}$.

The computation of errors is based on a standard jackknife technique. The IllustrisTNG300 box at $z=0$ is divided in 8 sub-boxes with $ L_{sub-box} = L_{box}/2 = 102.5$ $h^{-1}$~Mpc. The relative bias of Equation~\ref{eq:bias} is measured in 8 different configurations of equal volume, obtained by subtracting one sub-box at a time. The errors on the measured relative bias correspond to the standard deviation of all individual configurations. 
  
The 2-point correlation function is measured using CORRFUNC \citep{corrfunc2017} in a range of scales between 5 and 12 $h^{-1}$Mpc. We choose this range of scales due to the higher signal-to-noise in the assembly bias detection. The maximum scale is set at 12 $h^{-1}$Mpc in order to avoid problems derived from the relatively small size of the simulation box.

\section{Results}
\label{sec:results}

The manifestation of secondary halo bias on the clustering and properties of the central galaxy population is addressed in this section. We use several galaxy properties to split the galaxy population at fixed halo mass and measure the relative bias between subsets, as described in Section~\ref{sec:methodology}. Before showing the results derived from this analysis, we will present the secondary halo bias effect measured from IllustrisTNG300 (which serves as the {\it{reference signal}}) and discuss the correlations between halo and galaxy properties. 

\subsection{Secondary halo bias}
\label{sec:hab}

\begin{figure}
\begin{center}
\includegraphics[width=0.97\columnwidth]{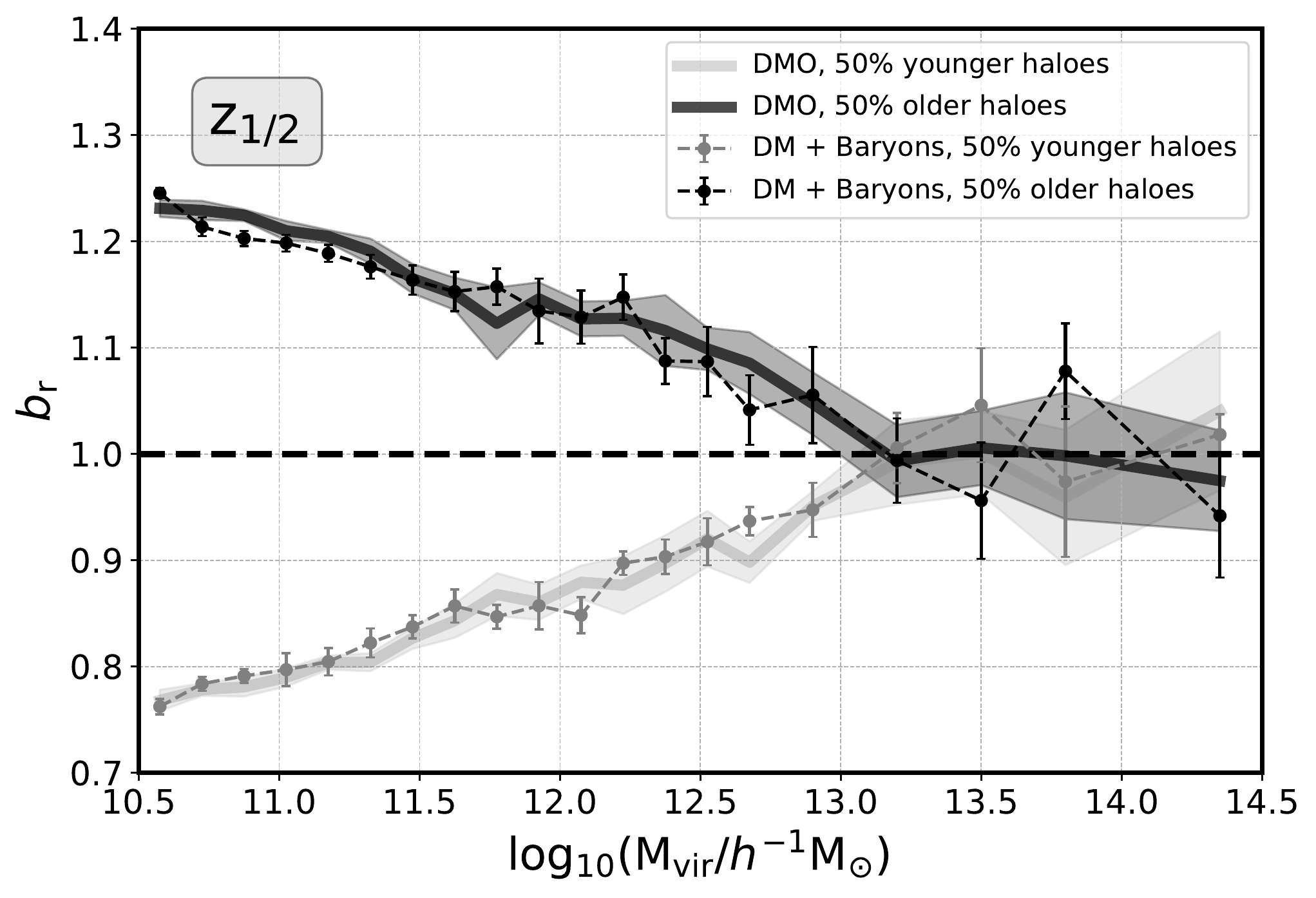}
\includegraphics[width=0.97\columnwidth]{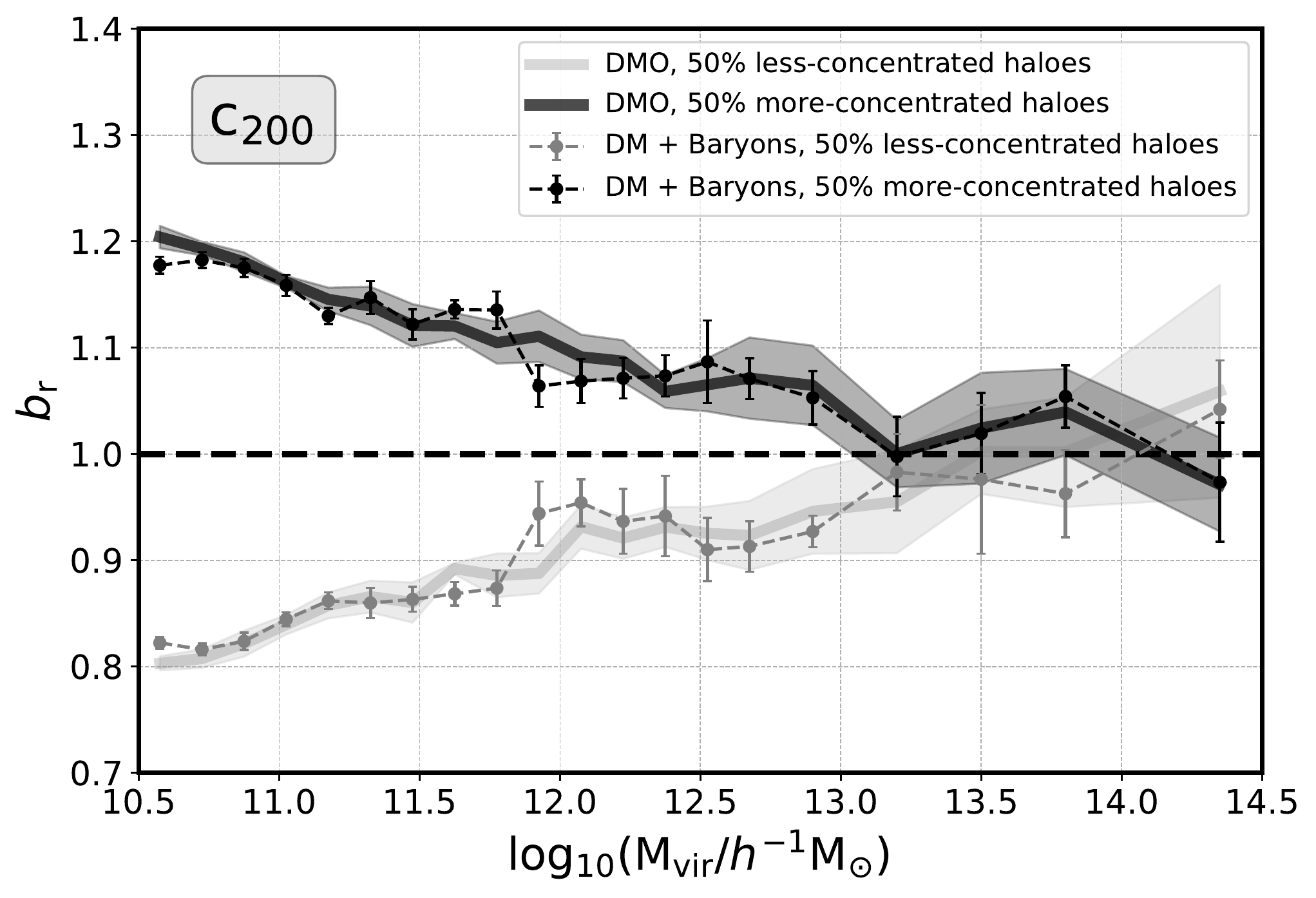}
\includegraphics[width=0.97\columnwidth]{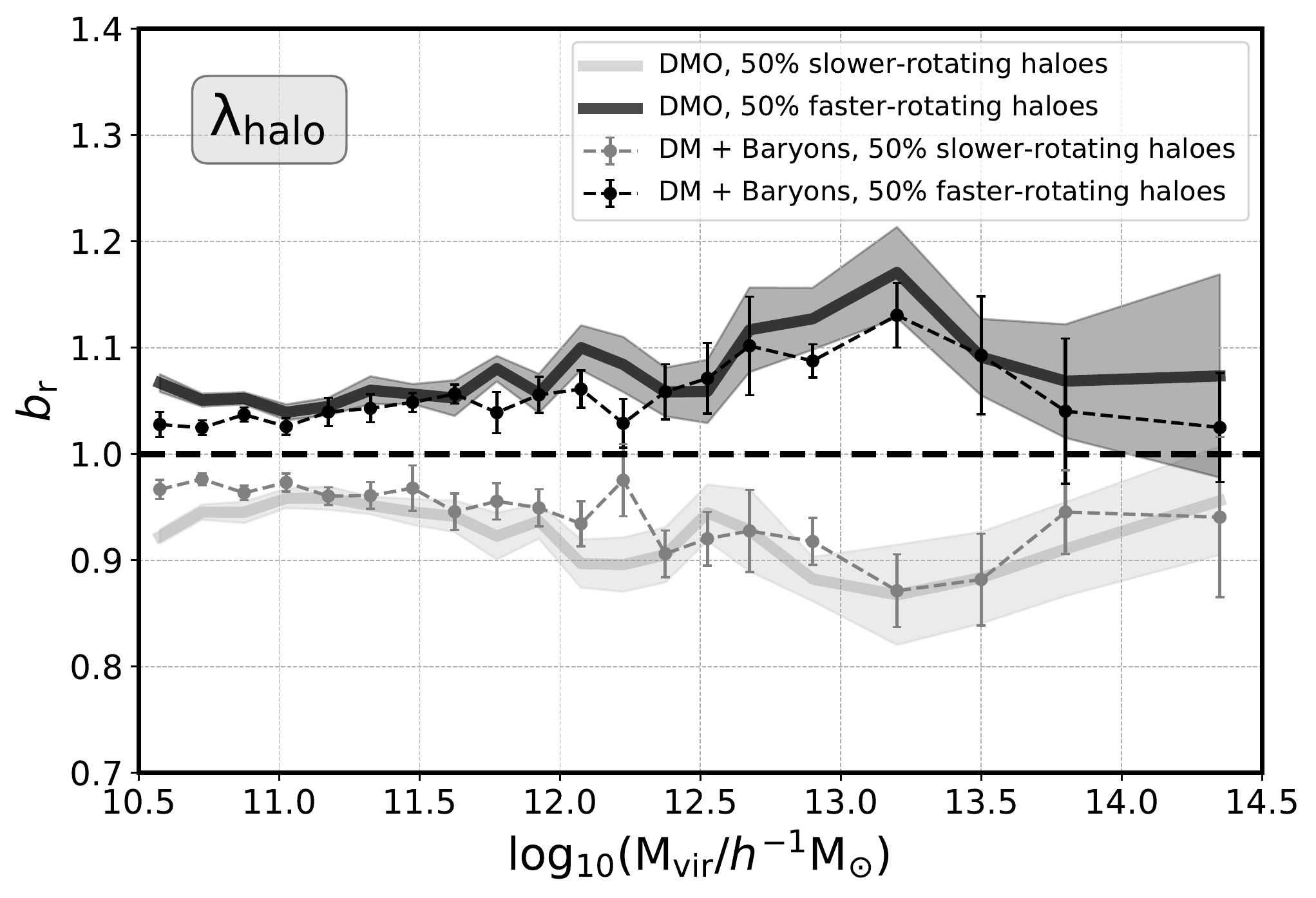}
\caption{Secondary halo bias for halo age (described by z$_{1/2}$), concentration (c$_{200}$) and spin ($\lambda_{\rm halo}$), for IllustrisTNG300 (dashed lines and dots) and IllustrisTNG300-DMO (solid lines) at $z=0$. In each panel, darker symbols/lines represent the 50$\%$ subset of haloes with higher values of the corresponding secondary property, whereas lighter symbols/lines show the remaining 50$\%$ lower-value subset.
Error bars display the jackknife error on the relative bias obtained from a set of 8 sub-boxes, as described in Section~\ref{sec:methodology}.}
\label{fig:hab}
\end{center}
\end{figure}

The dependence of halo clustering on multiple secondary halo properties at fixed halo mass (or velocity) has been measured extensively from different suites of N-body numerical simulations (see, e.g., \citealt{gao2005,wechsler2006,Gao2007,salcedo2018,han2018,SatoPolito2019}). As a sanity check, we begin by measuring the effect from haloes in both the IllustrisTNG300 and IllustrisTNG300-DMO boxes. From this comparison, the effect of baryons on the measured secondary bias effect can be assessed. 

Figure~\ref{fig:hab} presents the secondary bias signal for age (i.e., formation redshift, z$_{1/2}$), concentration, and spin. We recall that in order to maximise signal-to-noise, subsets of 50$\%$ of the entire distribution in each halo-mass bin are employed (instead of the standard 25$\%$ quartiles). We have checked that this choice has little effect, qualitatively, on the main conclusions of this paper. 

In order to further reduce the noise of the measurement at the high-mass end, where haloes are scarce, a varying bin size is adopted (see also \citealt{Xu2018}). At $\log_{10} ({\rm M_{\rm vir}}/h^{-1} {\rm M_{\odot}}) < 12.75$ we use $\Delta \log_{10}({\rm M_{vir}}) = 0.15$ (the typical value in these analyses), but we choose $\Delta \log_{10}({\rm M_{\rm vir}}) = 0.3$ above (with a larger single bin for very-high-mass haloes). 

For concentration and age, Figure~\ref{fig:hab} displays very similar trends. Older/more-concentrated haloes are more tightly clustered than younger/less-concentrated haloes below M$_{vir}\sim 10^{13}$ $h^{-1}$ M$_{\odot}$, as expected (with the effect being slightly larger for halo age). Above this halo mass, the trend appears to invert. Interestingly, this behaviour, the inversion of the signal, is very well documented for concentration, but not for z$_{1/2}$, for which most measurements show the signal vanishing (this is only strictly true for z$_{1/2}$, since \citealt{Chue2018} show that the 
amount of assembly bias at the high-mass end depends on the fraction of already formed mass used to define halo age). These are, however, the mass ranges for which we must be particularly careful, due to the low-number statistics of the sample. 

Also interesting in Figure~\ref{fig:hab} is the result for so-called {\it{spin bias}}, i.e., the secondary dependence on spin. Throughout the entire 
mass range, faster-rotating haloes are more tightly clustered than lower-spin haloes. This result is in some degree of tension with recent findings from N-body numerical simulations such as MultiDark or Vishnu (\citealt{SatoPolito2019} and \citealt{Johnson2019}, respectively), which show a statistically significant inversion of the trend at the low mass end (i.e. ``the spin-bias inversion"). Interestingly, as we discuss in Section~\ref{sec:spin} below, the crossover is indeed found when the selection is performed on the basis of the central galaxy spin. These apparent inconsistencies deserve further investigation, since they can potentially reveal clues on the origins of this secondary bias effect itself (Tucci et al. in prep.)\footnote{Note that in all previous works Rockstar was used to identify haloes, whereas IllustrisTNG300 employs a FOF algorithm. The {\it{spin-bias inversion}} was found for haloes of $\log_{10} ({\rm M_{vir}}/ h^{-1} {\rm M_{\odot}}) \simeq 11.5$, a range of masses where small differences in the halo definition could potentially have significant effects on the clustering measurements.}.

The scope of the present analysis is to show how secondary halo bias transmits to the galaxy population, and thus we will not concentrate on the particular details of the halo bias signal. We note also that the apparent decrease in the spin bias signal at the very high-mass end is probably an artefact due to the low number of haloes.

Figure~\ref{fig:hab} also shows that the effect of baryons on the secondary bias signal is not significant (the solid lines in each panel display the measurement in the DMO box). As expected, the trends are qualitatively very similar, with only small fluctuations that are consistent with the measured uncertainties.

\begin{figure*}
\includegraphics[width=0.97\columnwidth]{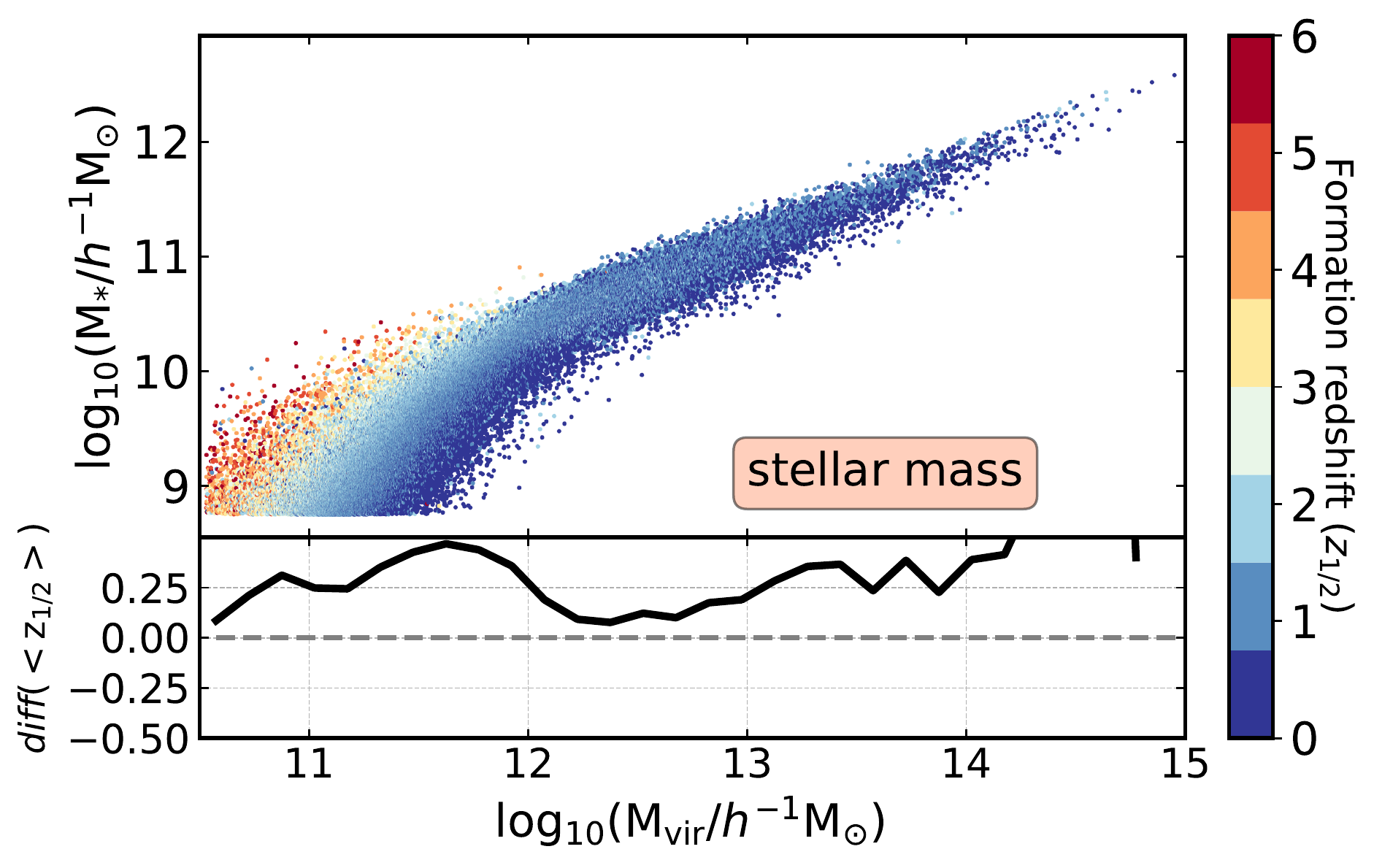}
\includegraphics[width=0.97\columnwidth]{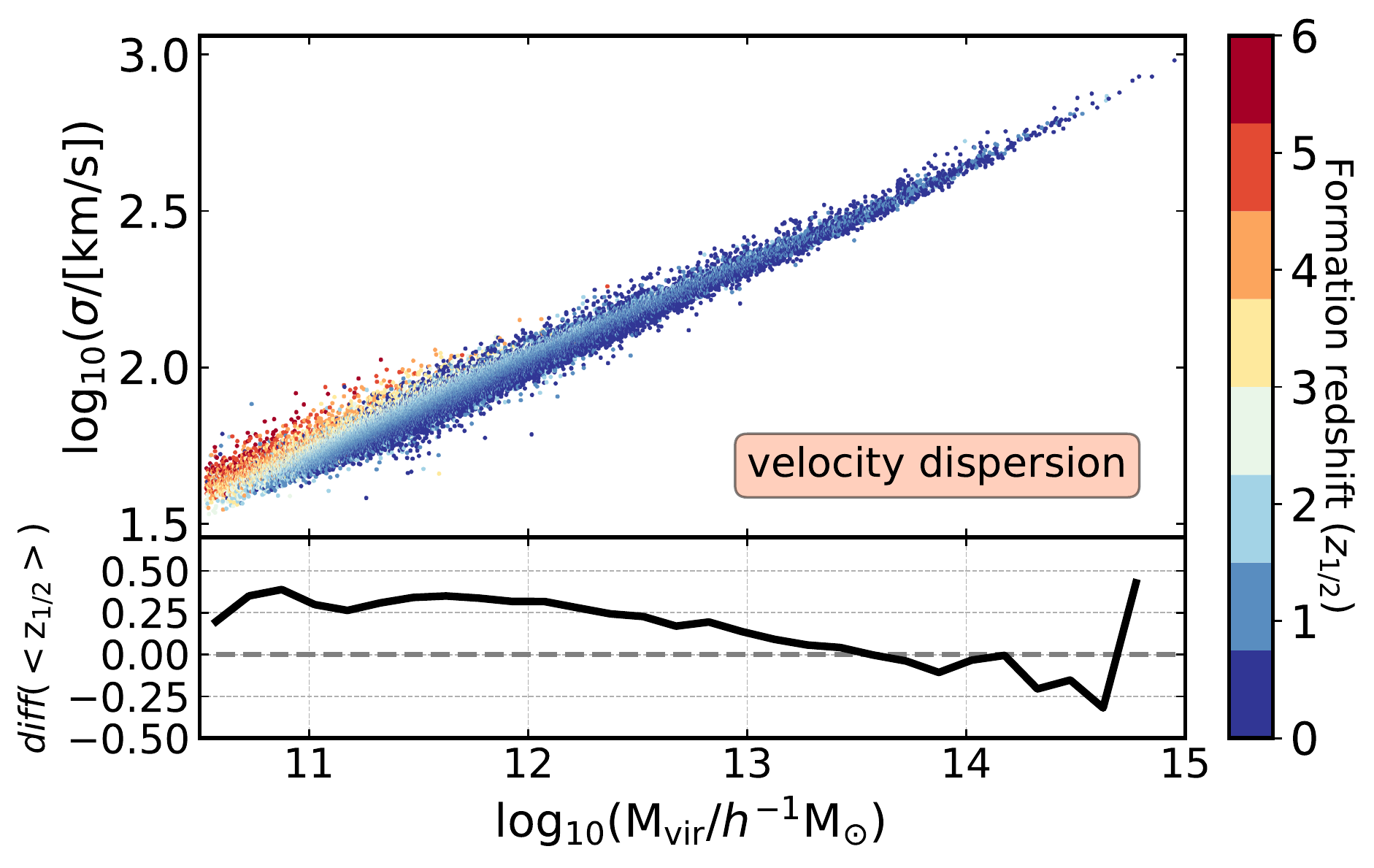}
\includegraphics[width=0.97\columnwidth]{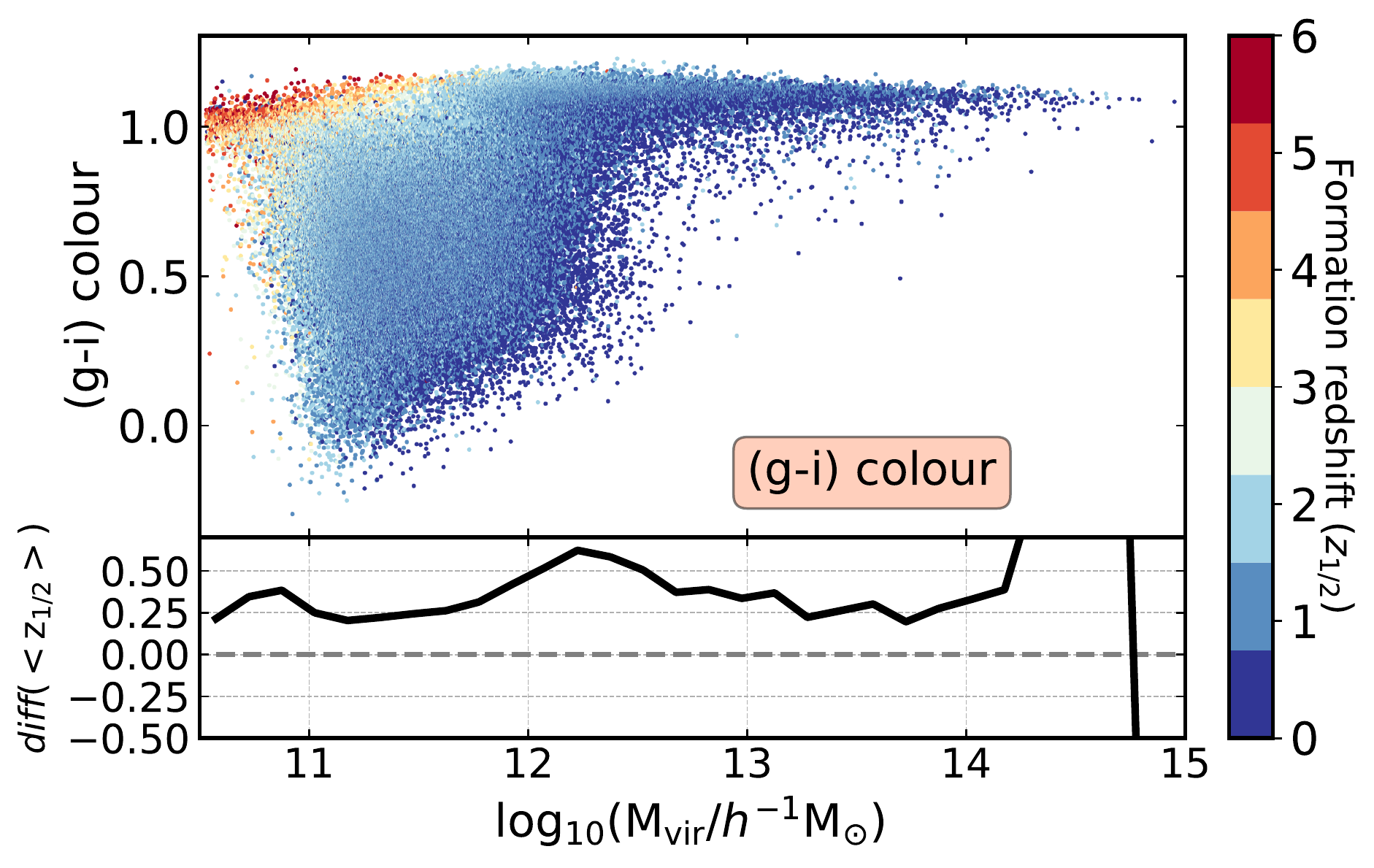}
\includegraphics[width=0.97\columnwidth]{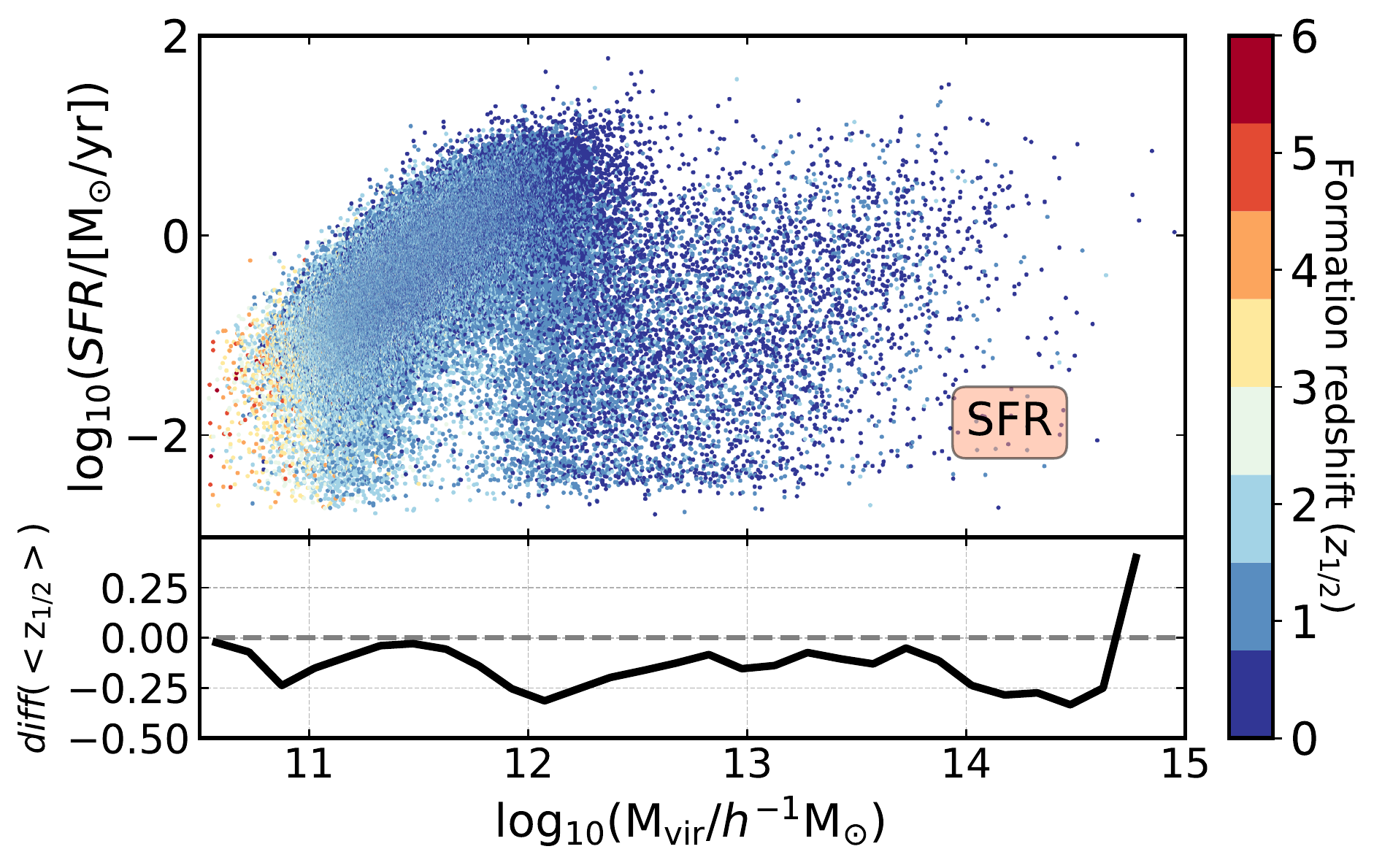}
\includegraphics[width=0.97\columnwidth]{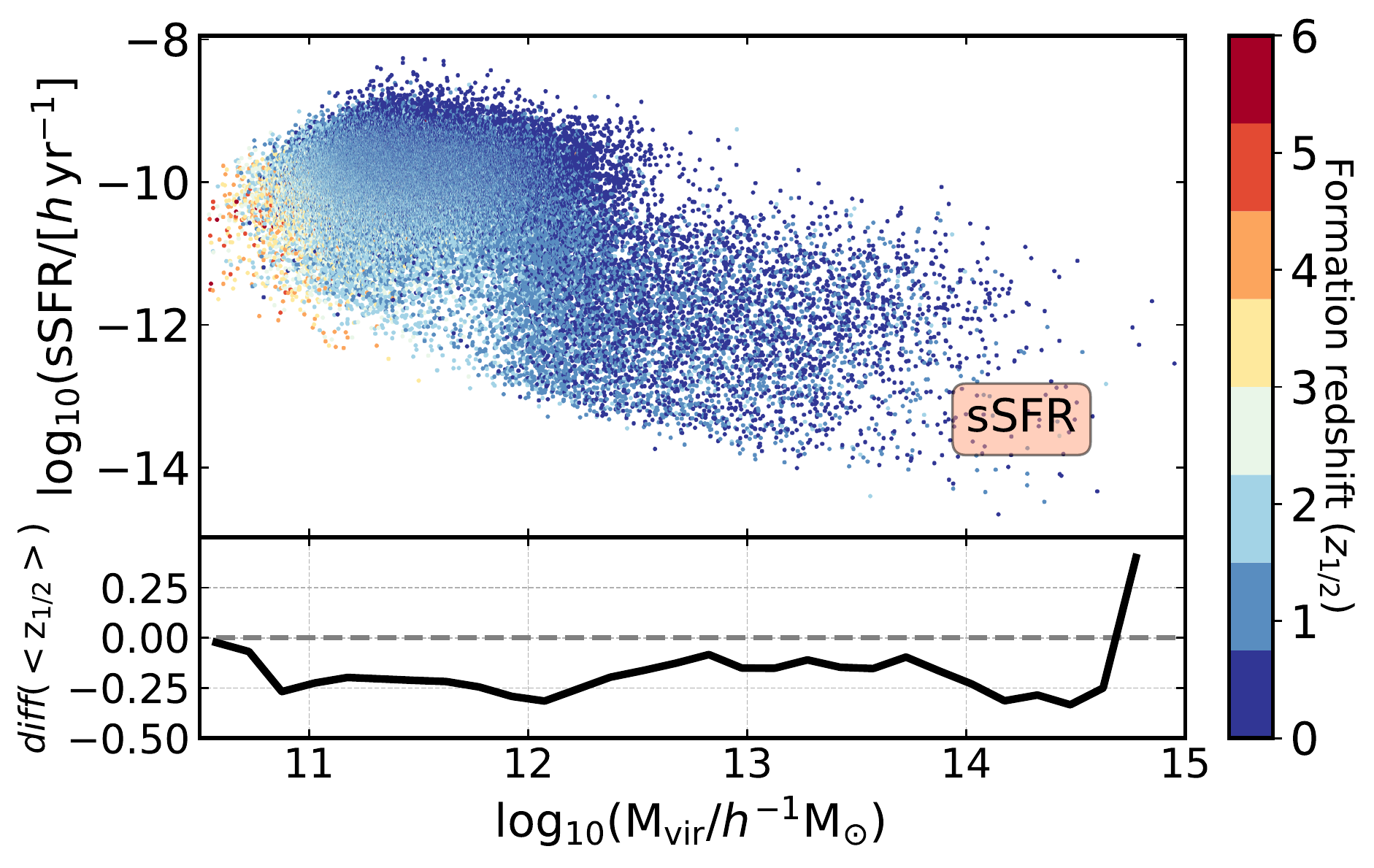}
\includegraphics[width=0.97\columnwidth]{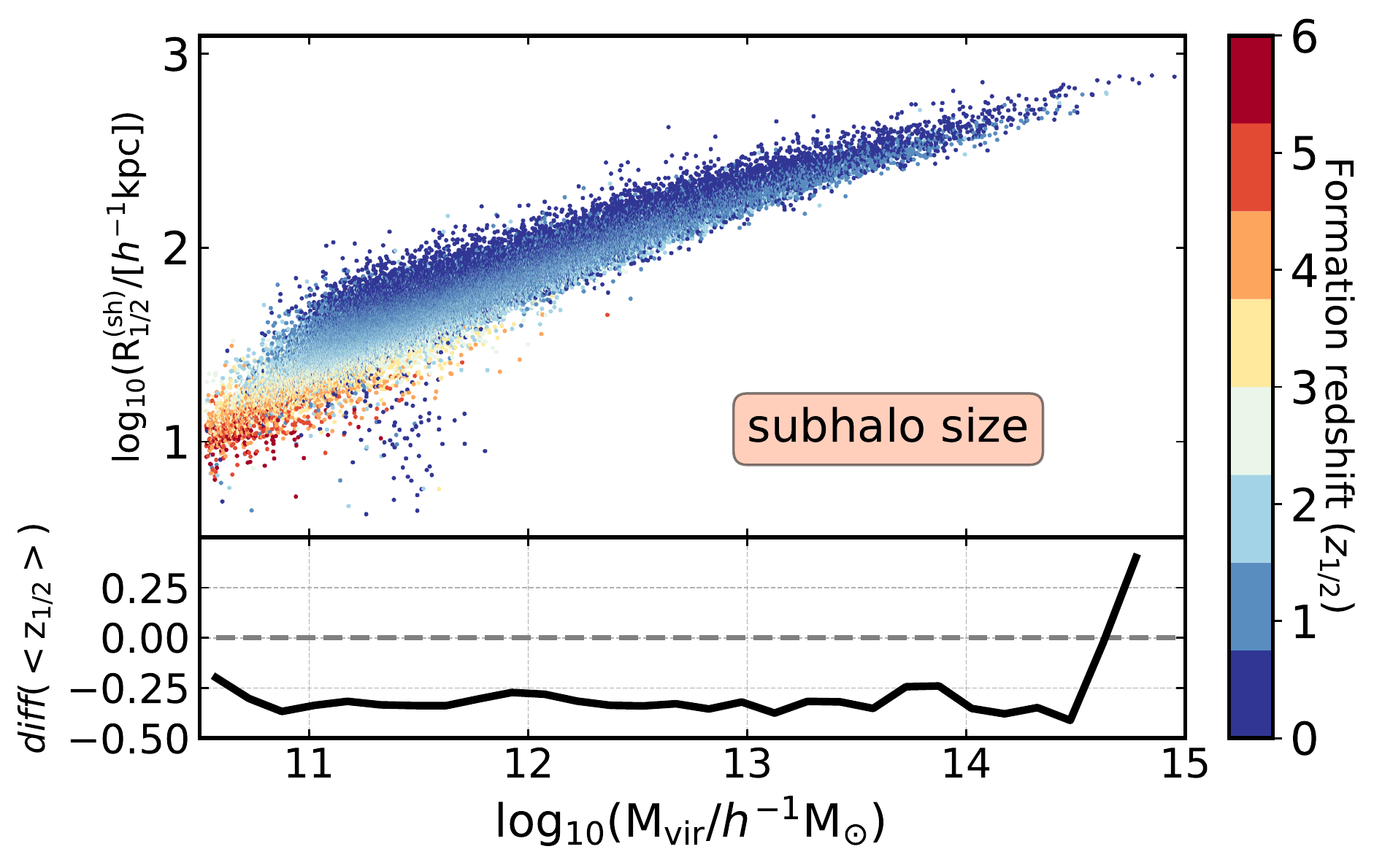}
\includegraphics[width=0.97\columnwidth]{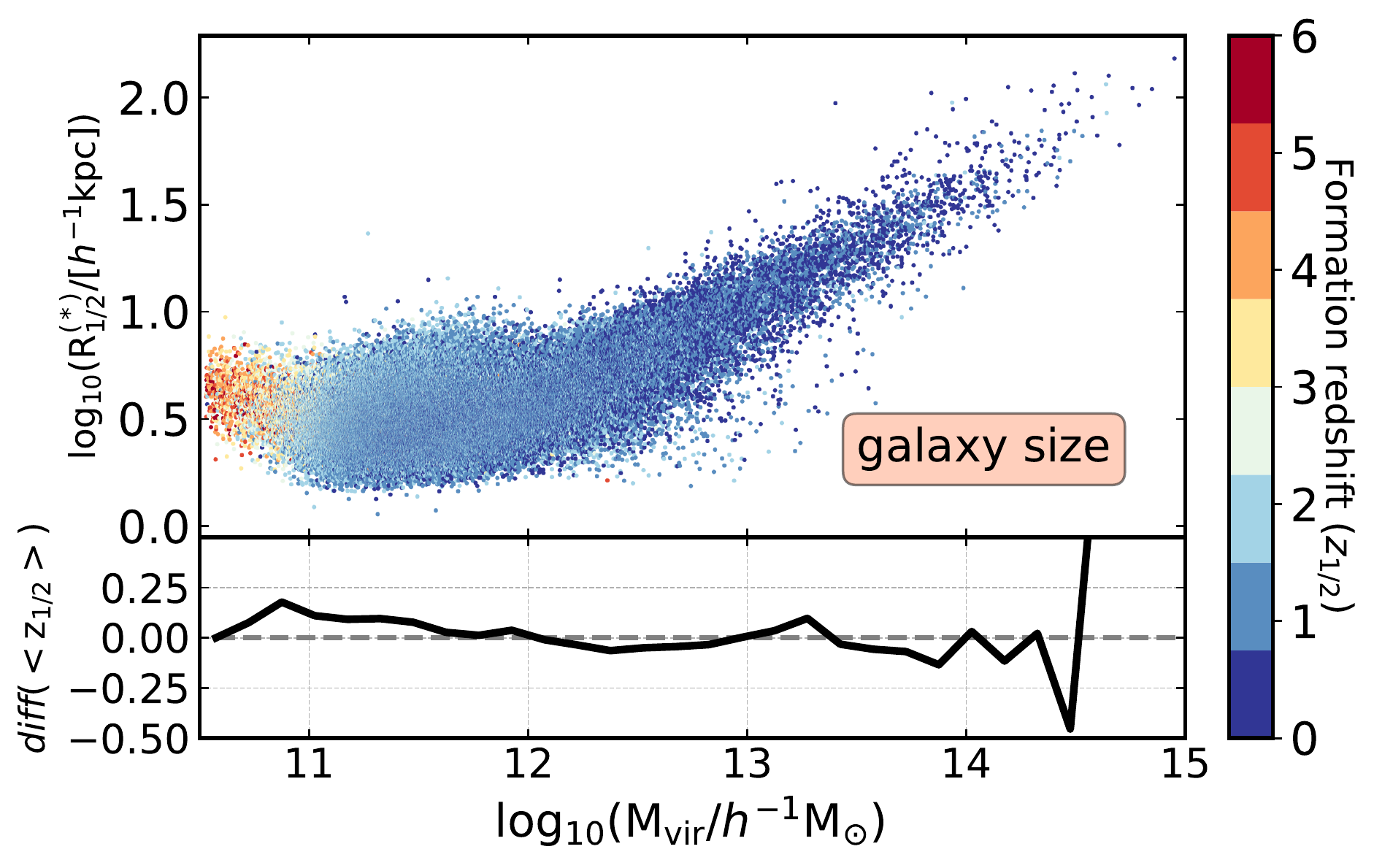}
\includegraphics[width=0.97\columnwidth]{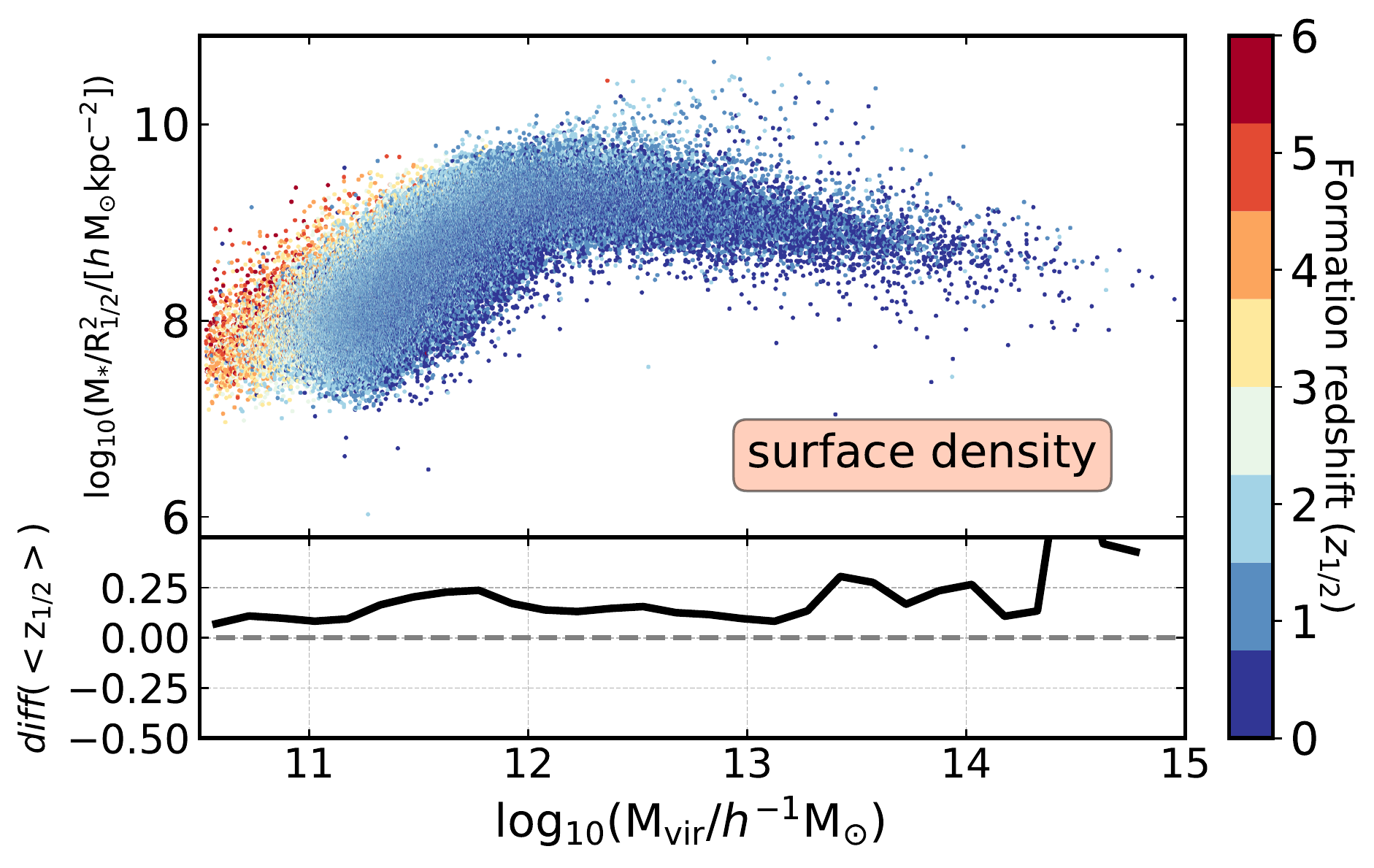}
\caption{The distributions of central galaxy properties as a function of halo mass. The colour code indicates the age of the halo where each galaxy lives, which is described by the halo formation redshift, z$_{1/2}$. In the subplots, the relative difference between the median z$_{1/2}$ 
in $50 \%$ subsets of the corresponding galaxy property is shown for reference, i.e. \textit{diff}$(<{\rm z}_{1/2}>) = \left(<{\rm z}_{1/2}^{(2)}> - <{\rm z}_{1/2}^{(1)}>\right)/<{\rm z}_{1/2}^{(1)}>$ (where 1 and 2 represent the bottom and top subsets, respectively).}
\label{fig:distributions}
\end{figure*}

\subsection{The correlations between halo and galaxy properties}
\label{sec:corr}

We analyse now the correlation between the properties of central galaxies and those of their hosting haloes. In Figure~\ref{fig:distributions}, central galaxy properties are displayed as a function of halo mass. Galaxies are also colour-coded according to halo formation time, z$_{1/2}$, which means that the halo assembly bias effect (i.e., the secondary bias on halo age) is completely characterised. In the subplots, the relative differences in the median z$_{1/2}$ in the top and bottom $50\%$ subsets of the corresponding galaxy property are shown for reference, i.e.  \textit{diff}$(<{\rm z}_{1/2}>) = \left(<{\rm z}_{1/2}^{(2)}> - <{\rm z}_{1/2}^{(1)}>\right)/<{\rm z}_{1/2}^{(1)}>$ (where 1 and 2 represent the bottom and top subsets, respectively). We focus here on stellar mass, velocity dispersion ($\sigma$), (g-i) colour, SFR, sSFR, subhalo half-mass radius R$_{1/2}^{(sh)}$, galaxy half-mass radius R$_{1/2}^{(*)}$, and surface density (see Section~\ref{sec:properties} for details).  

The following conclusions are drawn from this exercise:

\begin{itemize}
	\item As expected, more massive central galaxies typically inhabit more massive haloes. In addition, older haloes at fixed halo mass typically harbour more massive galaxies. Above $\sim$ 10$^{12}$ - 10$^{12.5}$ $h^{-1} {\rm M_{\odot}}$, the SHMR is flatter and exhibits less scatter. Also, the dependence on halo formation redshift at fixed halo mass appears to weaken \citep[see also,][]{Mathee2017}. Note that this effect is noticeable in the range of ages (colours) displayed as a function of halo mass, but it gets diluted in the fractional differences.
	
	\item The total velocity dispersion $\sigma$ (which takes into account all matter particles in the subhalo) is clearly the property that more tightly correlates with halo mass across the entire mass range. This is a reflection of DM being the dominant component. At fixed halo mass below M$_{\rm vir}$ $\lesssim 10^{13}$ $h^{-1}$ M$_{\odot}$, galaxies with higher $\sigma$ tend to reside in older haloes.  
	
	\item In lower-mass haloes (below the characteristic halo mass mentioned above), central galaxies exhibit larger scatter in colours, whereas higher-mass haloes are only inhabited by very red galaxies. At fixed halo mass, older haloes tend to host redder galaxies, but, again, the correlation seems stronger at the low-mass end. 
	
	\item Below the characteristic halo mass, there is a clear correlation (anti-correlation) between SFR (sSFR) and halo mass. Above this halo mass, the correlation appears weaker. The large scatter in the SFR for haloes above $\sim 10^{12.5} h^{-1} {\rm M_{\odot}}$ might be explained by the effect of AGN feedback, which prevents star formation in massive systems (this is also noticeable in the galaxy colours). The dependence on halo age at fixed halo mass is slightly more noticeable for sSFR than for SFR. 
	
	\item There is a tight correlation between the size of the baryonic + DM component of central galaxies (i.e. subhalo size) and the mass of the halo, with larger systems living in more massive haloes. At fixed halo mass, larger systems tend to prefer younger haloes. The strong correlation is again expected since subhalo size is mostly determined by the DM component and not by the baryonic component.
		
	\item When only the stellar component is considered (what is mostly measured in observations, i.e., R$_{1/2}^{(*)}$), the behaviour changes significantly. Galaxy size shows little dependence on halo mass for galaxies in haloes of ${\rm M_{vir}} \sim 10^{10.5} - 10^{12.5} ~h^{-1} {\rm M_\odot}$, while it increases with halo mass above $10^{12.5} ~h^{-1} {\rm M_\odot}$. These results are in agreement with those presented by \citet{Genel2018}. At fixed halo mass, there is almost no dependence on halo formation time.
	
	\item The surface (stellar) mass density of galaxies increases with halo mass below the characteristic halo mass range, but it remains fairly constant above. The dependence on halo formation time shows up again due to the inclusion of stellar mass. 
	
\end{itemize}

The different panels of Figure~\ref{fig:distributions} illustrate the importance of the range of halo masses around $\sim$ 10$^{12}$ -- 10$^{12.5}$ $h^{-1} {\rm M_{\odot}}$. Above this mass range, the efficiency in the production of stellar mass decreases (see the SHMR in the first panel of Figure~\ref{fig:distributions}, and, e.g., \citealt{RodriguezPuebla2015} for comparison), which transmits to many other halo--galaxy relations. This loss of efficiency is directly related to AGN feedback, which in these haloes becomes dominant with respect to other less-effective feedback mechanisms such as stellar feedback. Note that ``halo quenching", a process in which gas inflows are shock-heated to virial temperature preventing the accreted gas from fueling star formation, might also play a role
in massive haloes \citep[see e.g.,][]{Zu2016b}. As expected, the relations for subhalo size and total velocity dispersion are not affected by these processes, since these quantities are mostly determined by the DM component. 

Another important question is which galaxy properties better trace halo mass. Among the stellar population properties considered, stellar mass emerges as the most efficient tracer, although the scatter is large at the low halo-mass end. When the DM component is included, both size and, especially, the total velocity dispersion are the properties that exhibit less scatter. Note that several studies point to the {\it{stellar}} velocity dispersion as the best tracer of halo mass (e.g., \citealt{Zahid2016, Zahid2019}). This motivates the use of the velocity dispersion function (VDF, e.g., \citealt{MonteroDorta2017A}) in the context of halo--galaxy connection models.    
 
It is also noteworthy that halo age typically varies diagonally across the plane formed by a given galaxy property and halo mass. Haloes of the same age but higher mass tend to harbour typically more massive and redder galaxies. In other words, there is no particular galaxy property that completely determines halo formation redshift. On the other hand, the galaxy size, R$_{1/2}^{(*)}$, separates from the rest of the properties, as it is the only one that exhibits no correlation with halo age at fixed virial mass.

\begin{figure*}
\includegraphics[width=0.89\columnwidth]{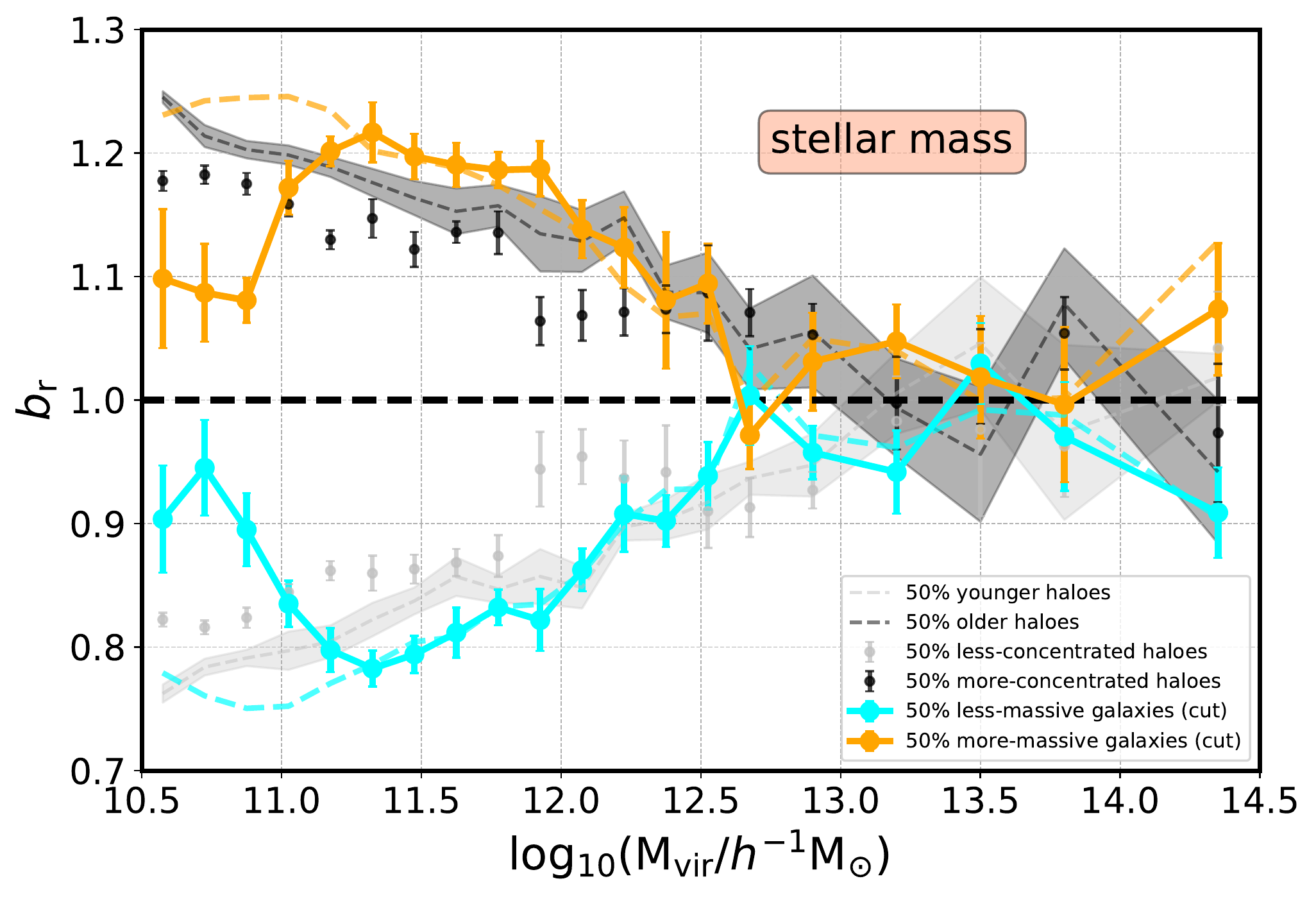}
\includegraphics[width=0.89\columnwidth]{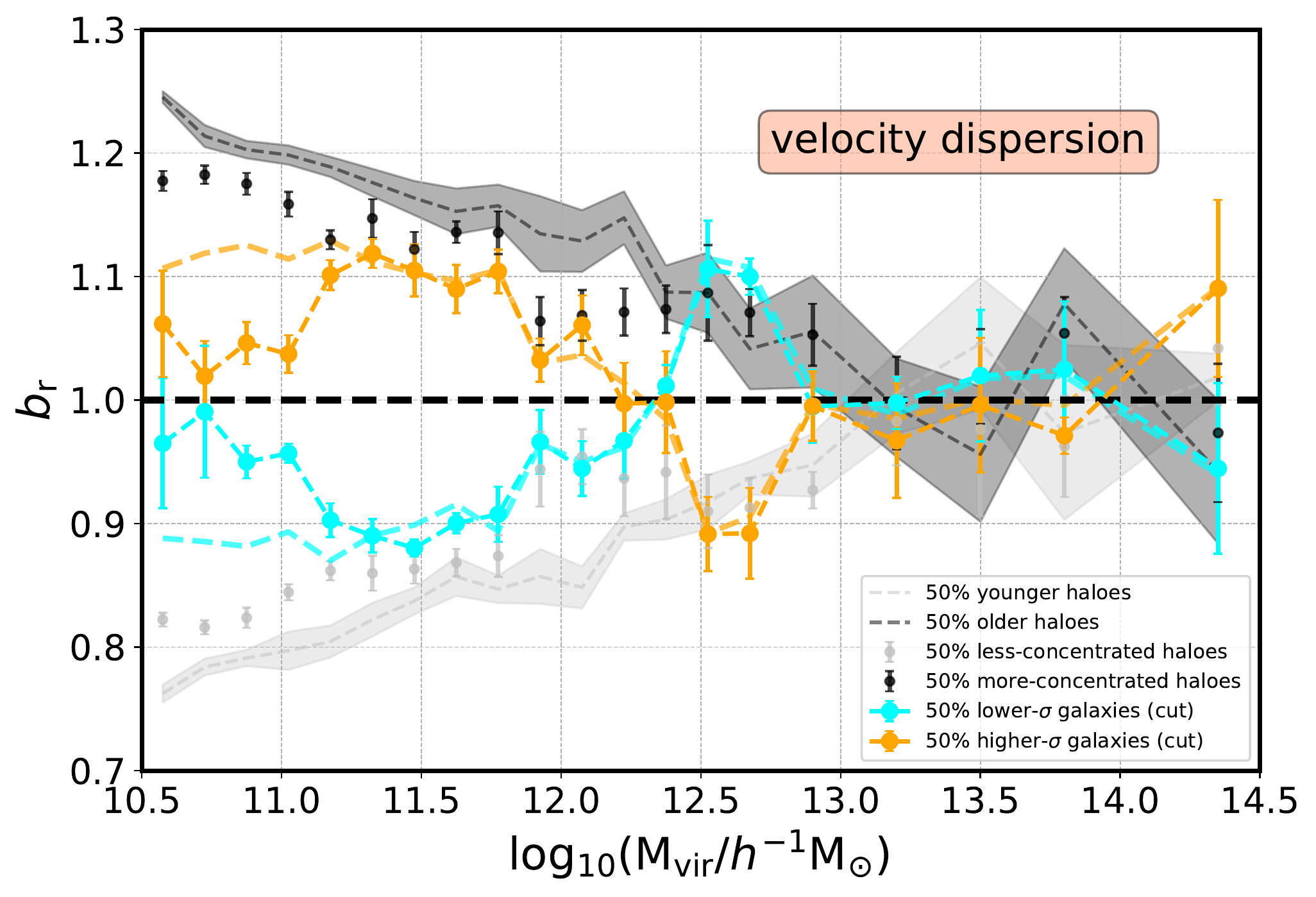}
\includegraphics[width=0.89\columnwidth]{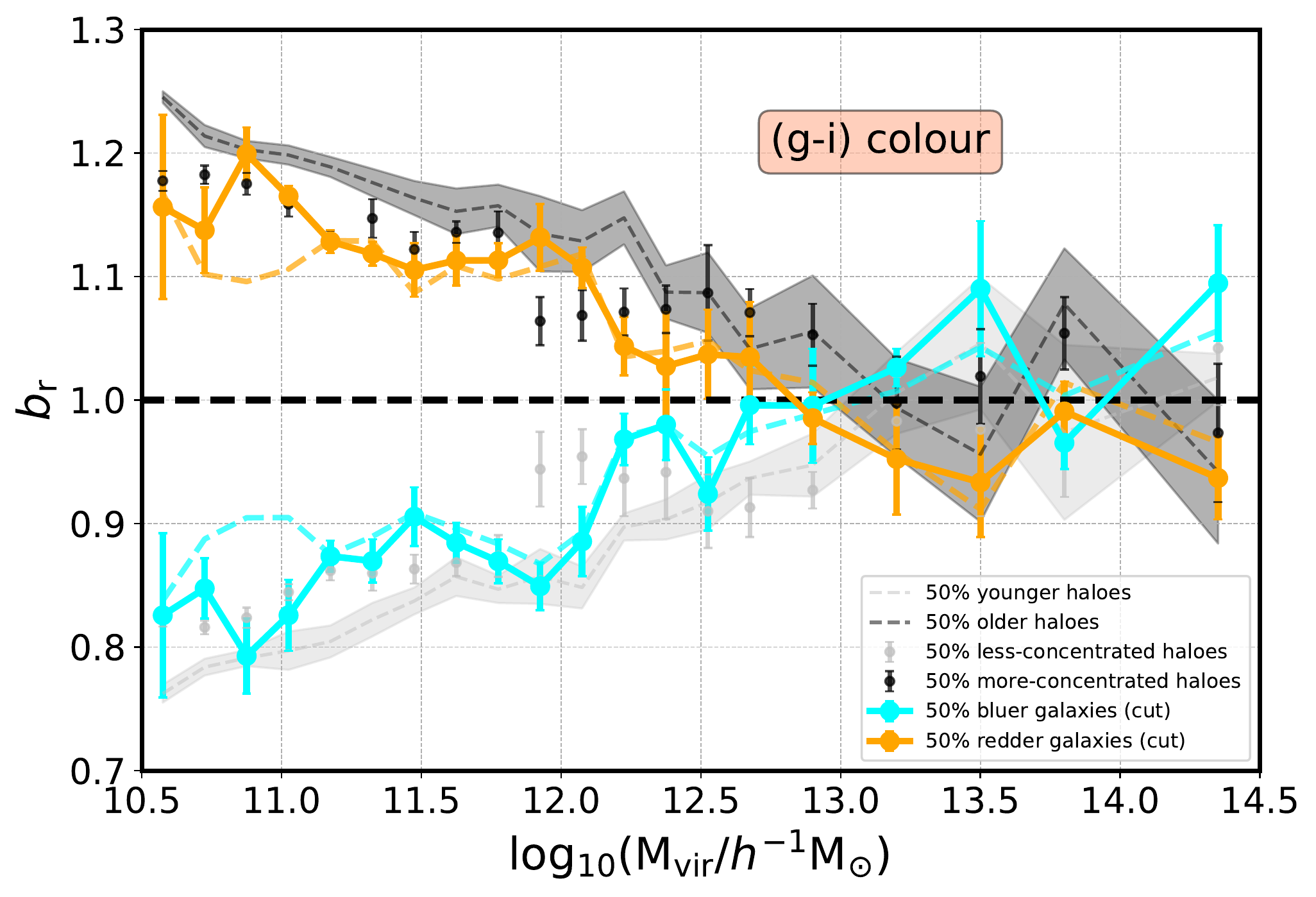}
\includegraphics[width=0.89\columnwidth]{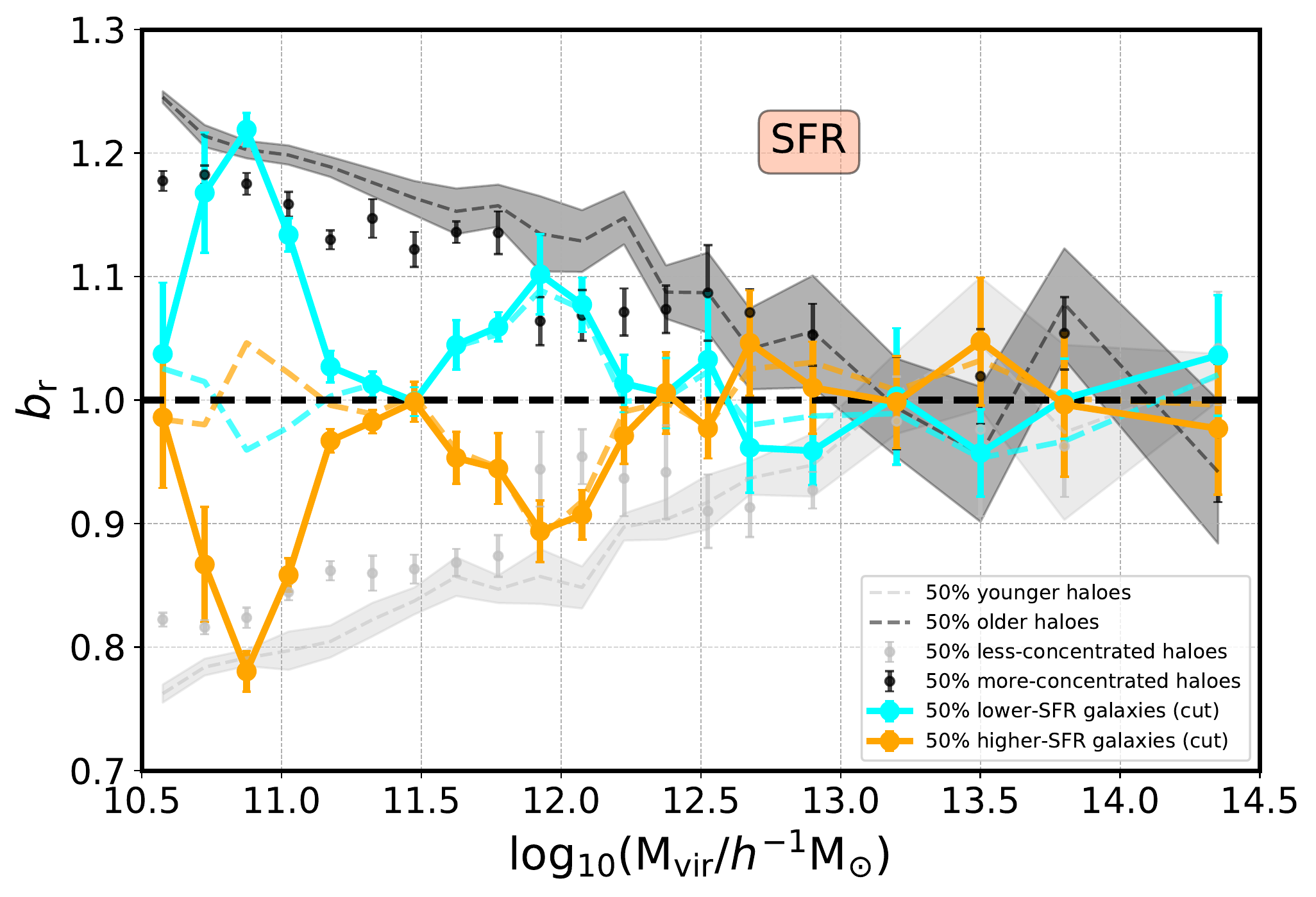}
\includegraphics[width=0.89\columnwidth]{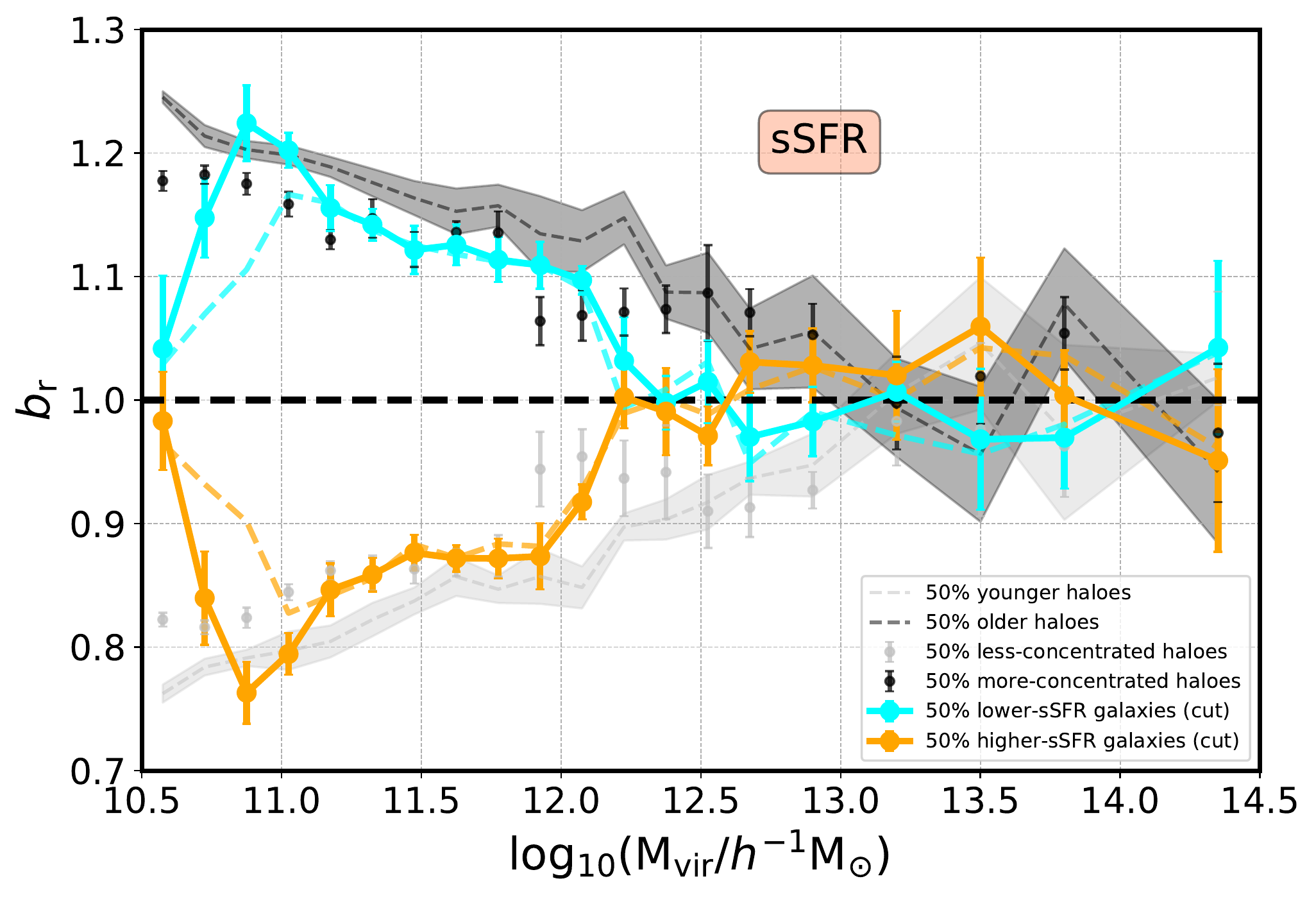}
\includegraphics[width=0.89\columnwidth]{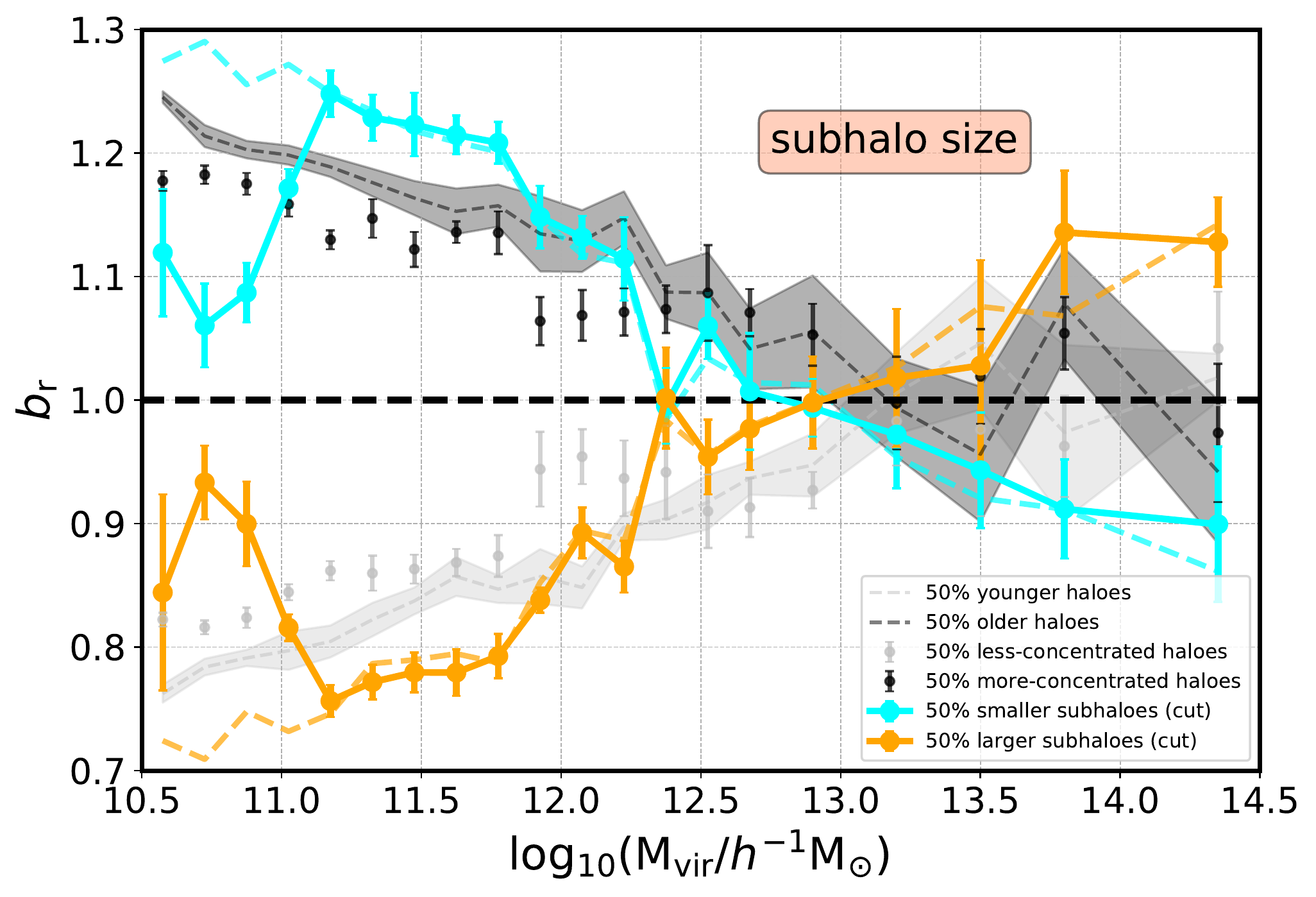}
\includegraphics[width=0.89\columnwidth]{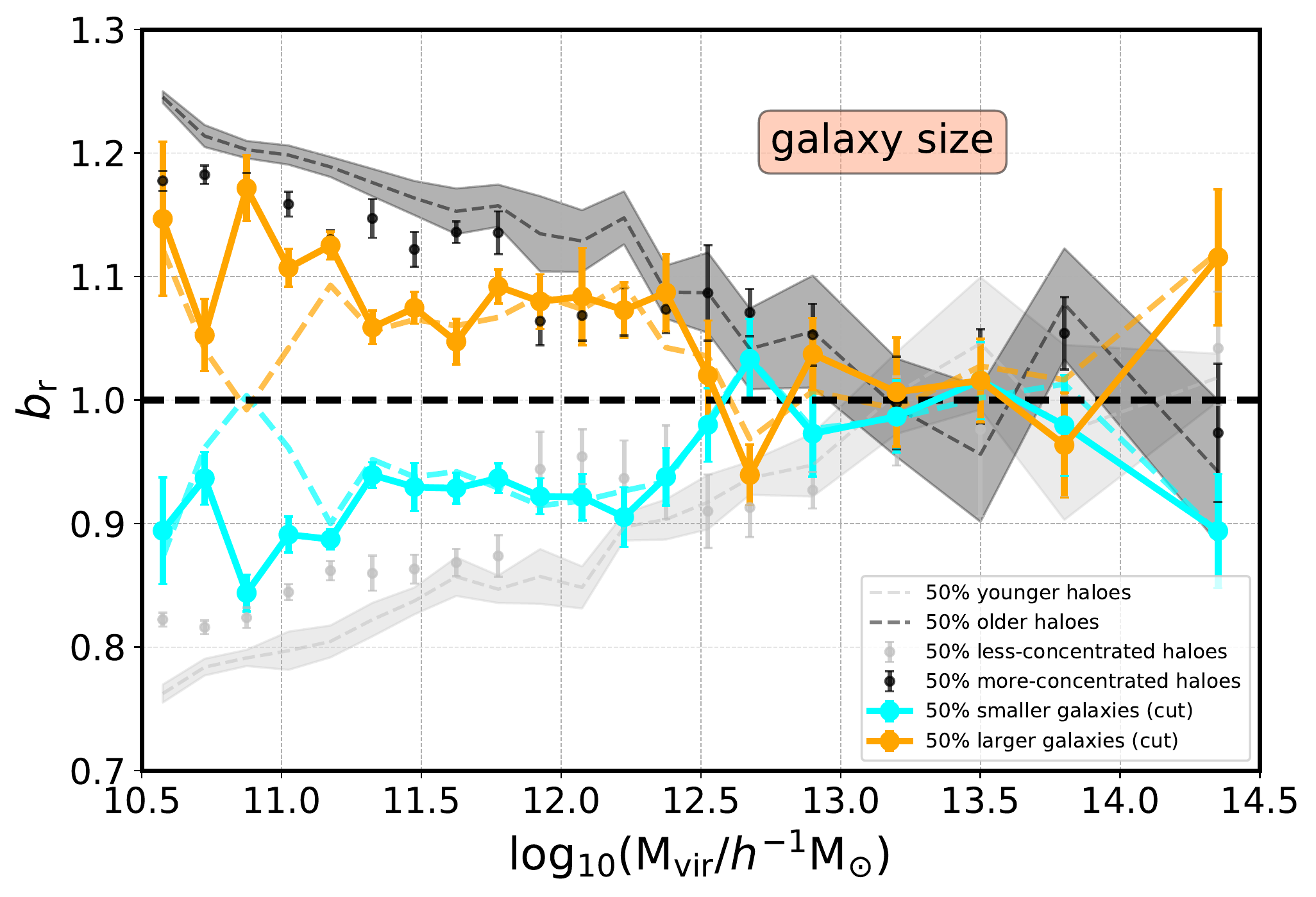}
\includegraphics[width=0.89\columnwidth]{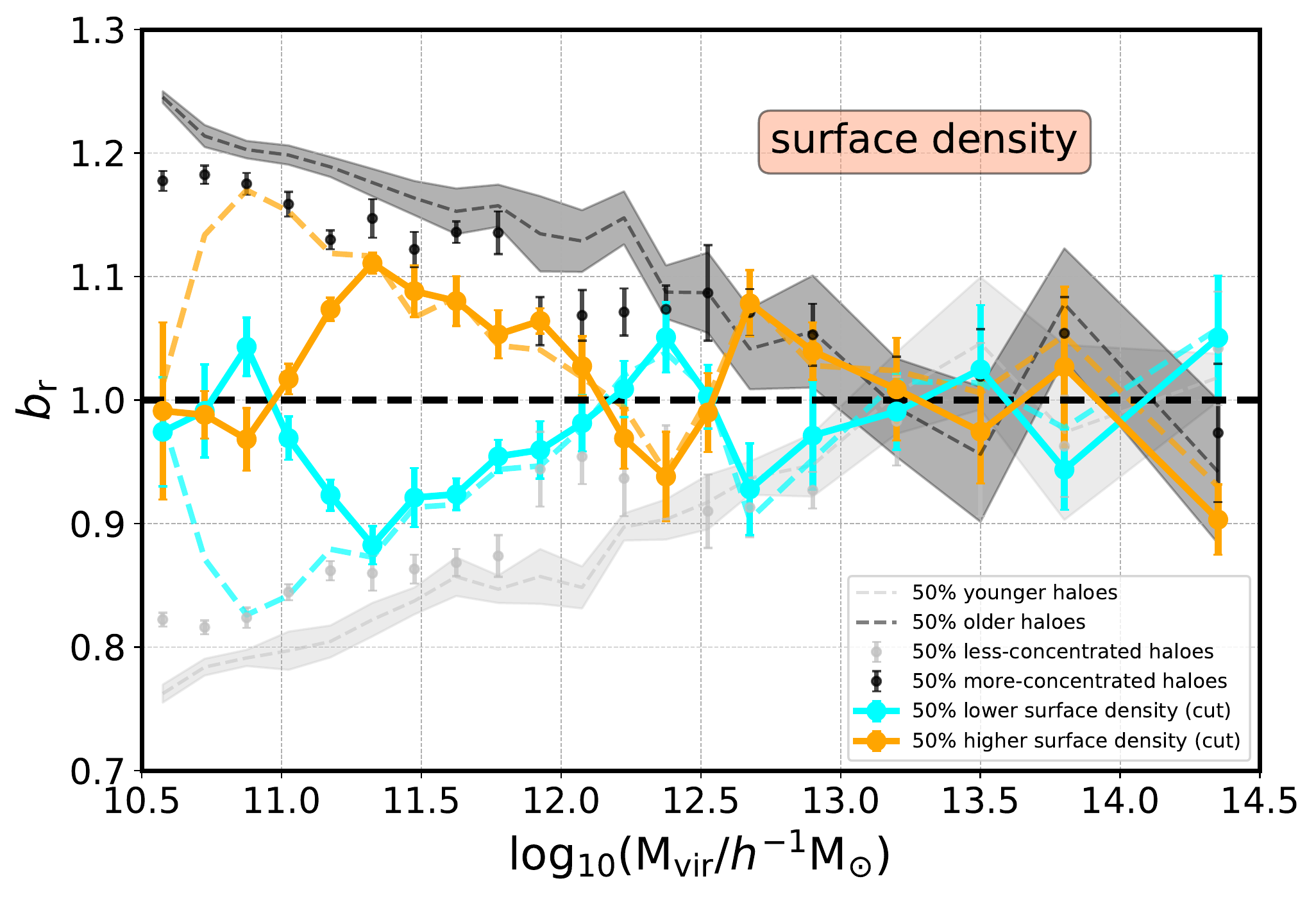}
\caption{The dependence of galaxy bias on several galaxy properties as a function
of halo mass. In each panel, orange dots represent the 50$\%$ subset of galaxies with higher values of the corresponding galaxy property, whereas cyan dots show the remaining 50$\%$ lower-value subset. These results are obtained assuming a stellar-mass cut of $\log_{10} ({\rm M_*}/ h^{-1} {\rm M_{\odot}}) > 8.75$, which implies a minimum subhalo-mass resolution of 50 particles. Orange/cyan dashed lines show the same measurements assuming no stellar-mass cut. For reference, the secondary halo bias results for concentration (black/grey dots with error bars) and formation redshift (black/grey dashed lines with error bands) are shown in the background. Errors bars/bands show the jackknife uncertainties obtained from a set of 8 equal-volume sub-boxes.}
\label{fig:gab1}
\end{figure*}

\begin{figure*}
\includegraphics[width=\columnwidth]{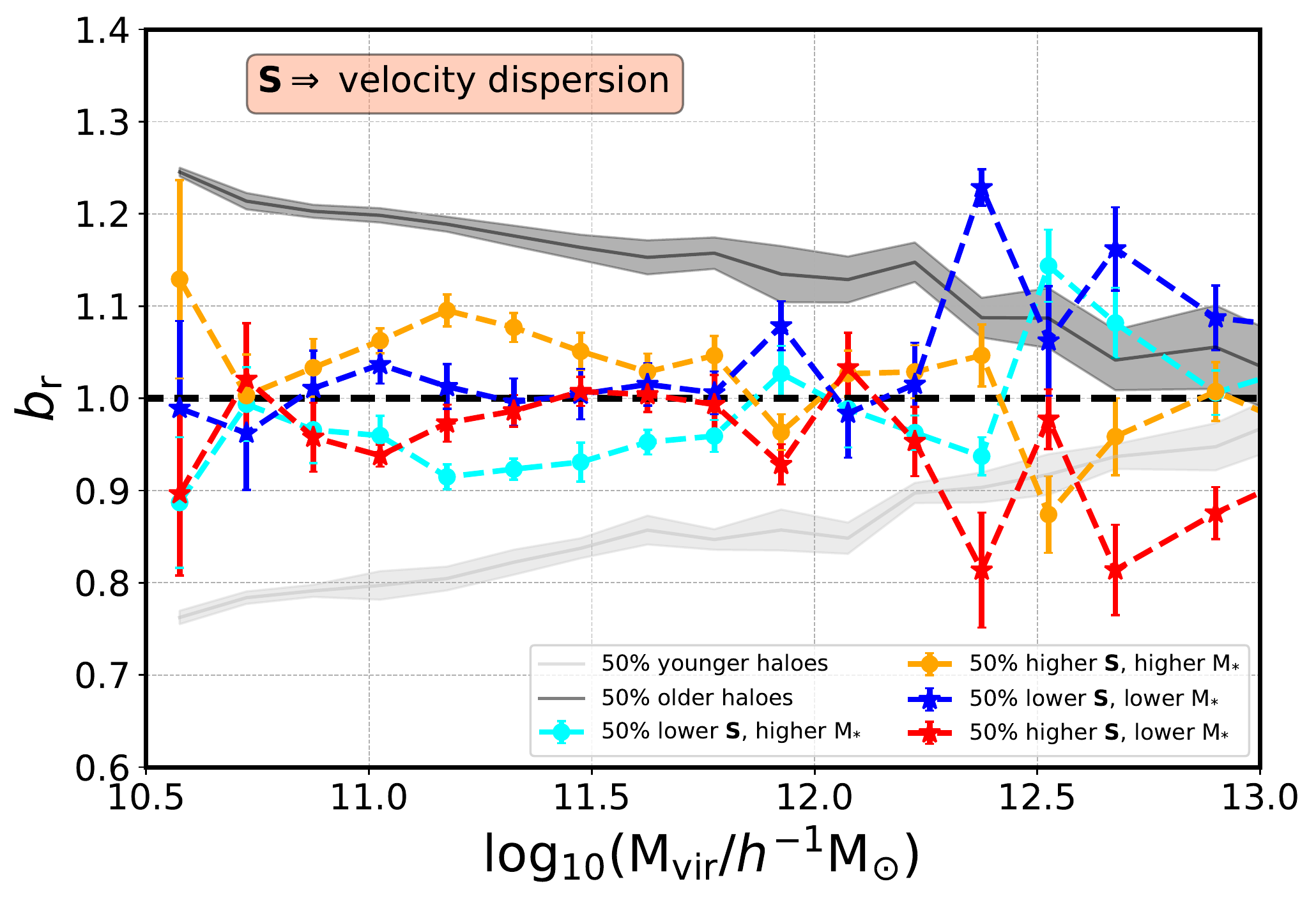}
\includegraphics[width=\columnwidth]{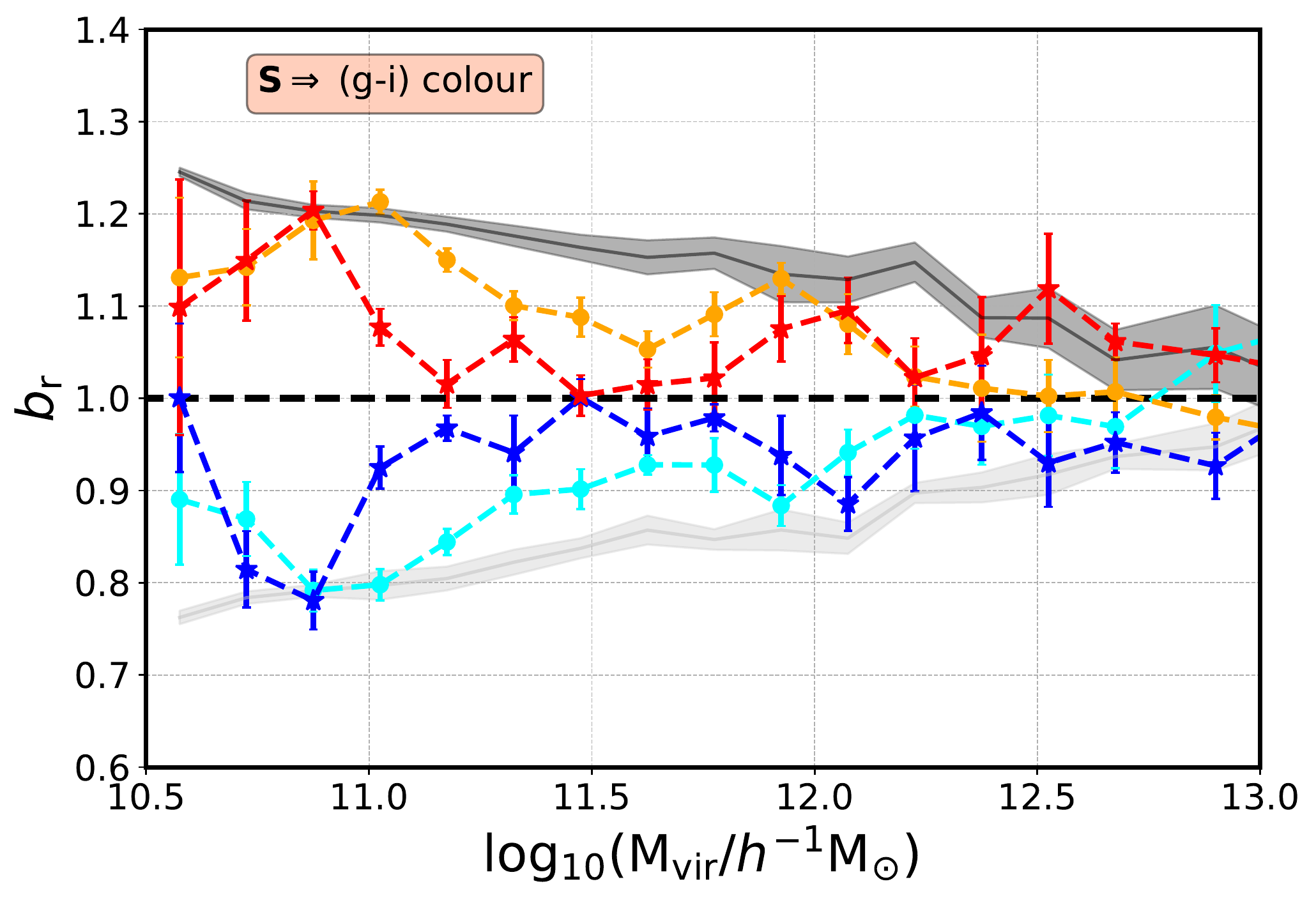}
\includegraphics[width=\columnwidth]{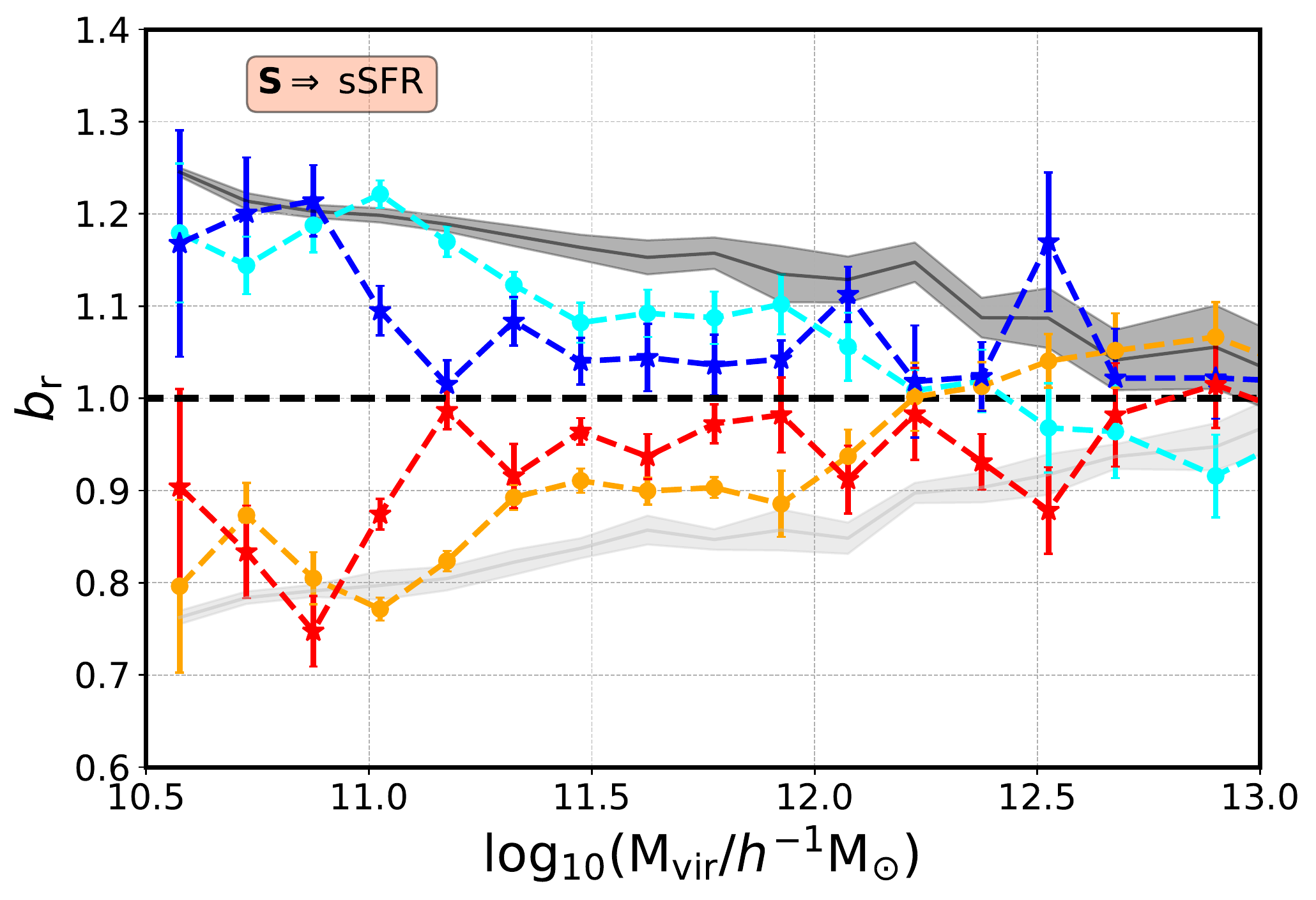}
\includegraphics[width=\columnwidth]{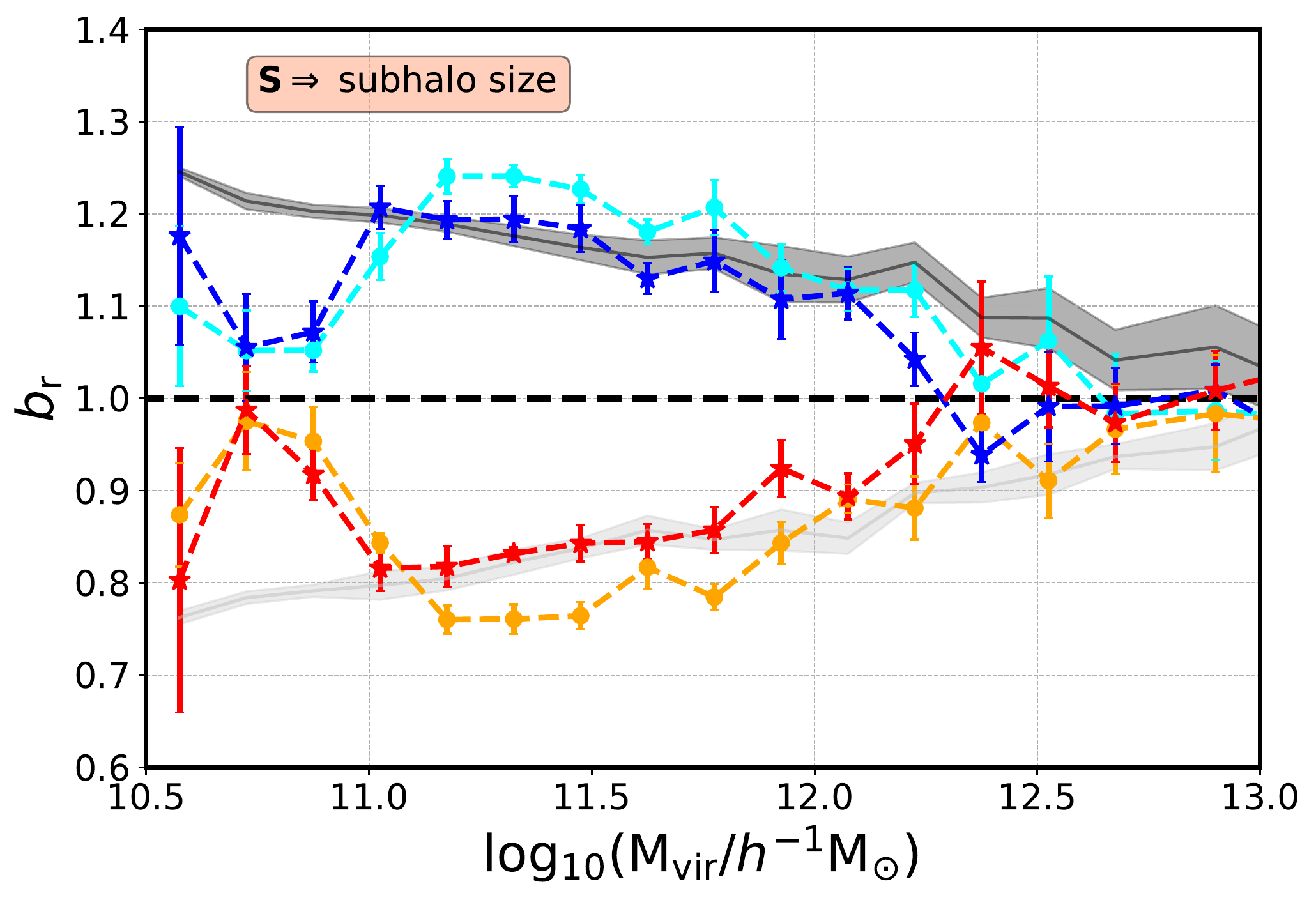}
\includegraphics[width=\columnwidth]{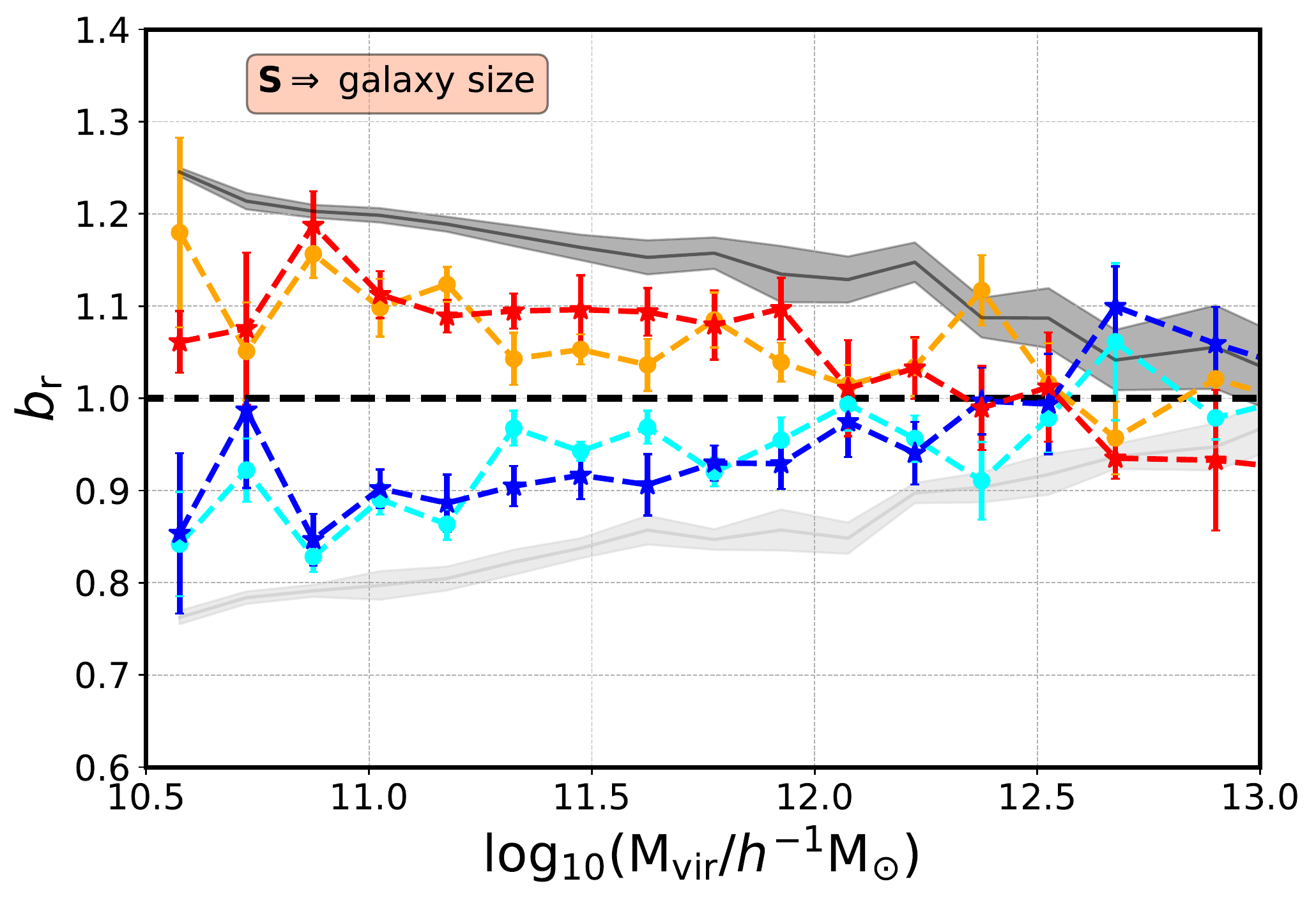}
\includegraphics[width=\columnwidth]{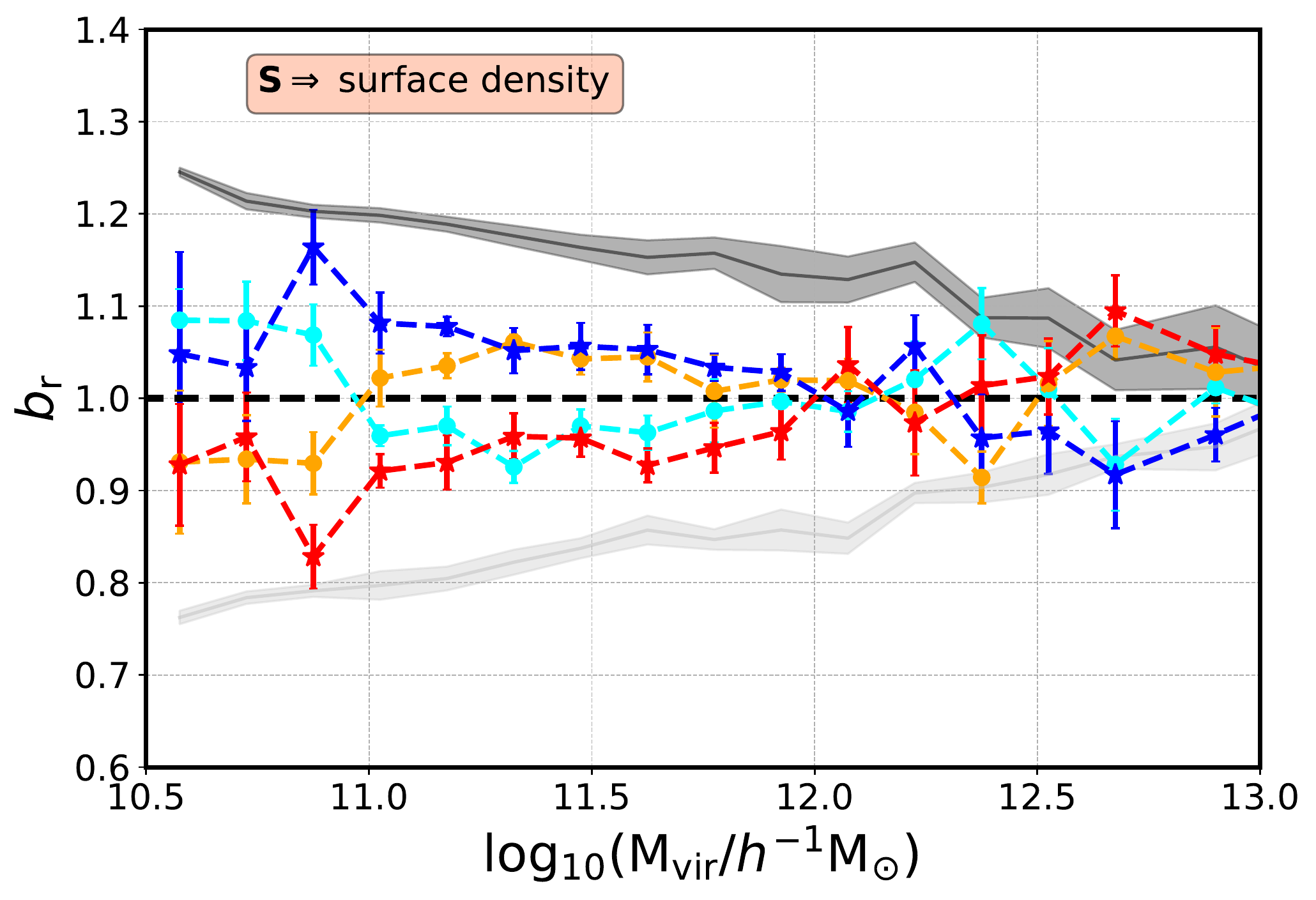}
\caption{The dependence of galaxy bias on several galaxy properties
as a function of halo mass and stellar mass. Orange/cyan symbols show the secondary galaxy bias signal for high-mass galaxies (50$\%$ of the distribution), whereas red/blue symbols show results for lower-mass galaxies. The secondary halo bias effect on halo age is kept in the background for reference (black/grey solid lines). The error bars/bands represent the jackknife uncertainties obtained from a set of 8 divisions.}
\label{fig:gab2}
\end{figure*}

\subsection{Galaxy clustering at fixed halo mass}
\label{sec:gabm}

The correlations and scatters between halo and galaxy properties along with the distribution of biases dictate how the secondary bias signal transmits to the central galaxy population when different galaxy selections are applied. In Figure~\ref{fig:gab1}, we select galaxies according to the same properties discussed in the previous sections, and measure relative galaxy bias at fixed halo mass as described in Section~\ref{sec:methodology}. For reference, Figure~\ref{fig:gab1} displays the secondary halo bias on concentration and age in the background (i.e., halo assembly bias) which are the halo secondary properties more likely to be connected to the galaxy properties used here (the possible manifestation of the spin bias signal will be addressed in Section~\ref{sec:spin}). 

From the very first glimpse at Figure~\ref{fig:gab1}, it becomes evident that the clustering of galaxies at fixed halo mass depends on multiple galaxy properties. Statistically-significant signals are detected for all properties (although the measurement is very noisy for SFR).
The comparison with the halo assembly bias signal on the background and the halo--galaxy relations of Figure~\ref{fig:distributions} clearly indicates that the galaxy clustering difference is, in most cases, due to the mapping between galaxy and halo properties, at fixed halo mass. These measurements thus represent a prediction of galaxy assembly bias, {\it{according to the definition adopted in this analysis}}. This galaxy assembly bias signal was already presented, albeit with less statistical significance, in \cite{Xu2018}, who used the 75-$h^{-1}$Mpc Illustris-2 box. As mentioned before, no conclusive, irrefutable evidence has been found with real data. Among the main issues that beset this measurement are: the large uncertainties in the determination of halo masses, the contamination from satellites, and even the uncertainties in stellar population synthesis (see \citealt{Miyatake2015,Lin2016,Sunayama2016,MonteroDorta2017B,Niemiec2018} for more discussion). 

Among the different galaxy properties featured in Figure~\ref{fig:gab1}, the stellar mass, (g-i) colour, sSFR, and subhalo size are the ones that more clearly follow the halo assembly bias signal (note that subhalo size is considered a galaxy property here for simplicity, but it is actually determined by the DM component). Weaker/noisier results are found for velocity dispersion, surface density, and especially SFR, although the dependence at fixed halo mass is still compatible with the background signal. 

The galaxy size emerges in Figure~\ref{fig:gab1} as a special case. While smaller subhaloes are more tightly clustered than larger subhaloes, this trend inverts when only the stellar component (galaxy size) is considered. Larger central galaxies are significantly more clustered than their smaller counterparts up to, at least, $\log_{10} ({\rm M_{vir}}/ h^{-1} {\rm M_{\odot}}) \sim 12.5$. Interestingly, Figure~\ref{fig:distributions} shows little to no correlation between galaxy size and halo age at fixed halo mass. This could indicate that the difference in clustering displayed in Figure~\ref{fig:gab1} is not due to halo assembly bias.

It seems clear that the scatter in the halo--galaxy connection does not wash out the assembly bias signal (as long as halo mass is known with precision). Another interesting question that Figure~\ref{fig:gab1} poses is whether the signal measured for several properties at fixed halo mass also depends on stellar mass. As shown in the upper left panel of Figure~\ref{fig:gab1}, more massive galaxies are more tightly clustered than less massive galaxies in haloes of similar mass. In order to investigate the stellar mass effect on galaxy assembly bias, we take advantage of the statistics provided by IllustrisTNG300 to additionally split the population in each halo mass bin by stellar mass. Figure~\ref{fig:gab2} displays the galaxy assembly bias signal for velocity dispersion, colour, sSFR, half-mass radius (using both definitions) and surface density for the 50$\%$ higher stellar mass sub-population (orange/cyan colours) and the 50$\%$ lower stellar mass sub-population (red/blue colours). 

Figure~\ref{fig:gab2} reveals that the implicit clustering dependence on stellar mass has different effects depending on the galaxy property used to perform the subset selection. Although results are unavoidably noisier than those presented in Figure~\ref{fig:gab1}, the following conclusions can be drawn, qualitatively, from this analysis: 

\begin{itemize}
	\item The galaxy assembly bias signal for properties such as colour, sSFR, and subhalo size displays a clear dependence on stellar mass: the signal is always stronger (i.e., larger clustering difference between S-subsets) for higher-mass galaxies than for lower-mass galaxies. Relative to the magnitude of the signal itself, the dependence is, as expected, weaker for subhalo size, which is very connected to the DM content of the halo.
	\item The effect of stellar mass seems insignificant for galaxy size: higher and lower-mass galaxies exhibits statistically similar assembly bias signals.
	\item Surface density shows an interesting inversion of the signal for lower-mass galaxies (this is also found at low statistical significance for velocity dispersion). For lower-mass galaxies, the relative bias is higher for objects with low surface density, with the opposite trend happening for higher-mass galaxies.
	\end{itemize}

Figure~\ref{fig:gab1} and~\ref{fig:gab2} show some particular cases where the galaxy assembly bias signal exceeds the halo assembly bias signal in some mass ranges (i.e., for stellar mass and subhalo size). Note that this does not necessarily imply any additional signal that only depends on galaxies (i.e., on the galaxy formation process). In a given halo mass bin, by choosing half the population by stellar mass, we could be essentially selecting slightly older/younger haloes than what a halo-age-based selection would produce (i.e., it depends on the function that maps stellar mass and age, at fixed halo mass). It is also possible that by selecting on the basis of these galaxy properties we are effectively mapping a combination of multiple secondary halo bias effects (e.g., age, concentration, spin, etcetera).

\subsection{The effect of spin bias}
\label{sec:spin}

\begin{figure}
\includegraphics[width=0.9\columnwidth]{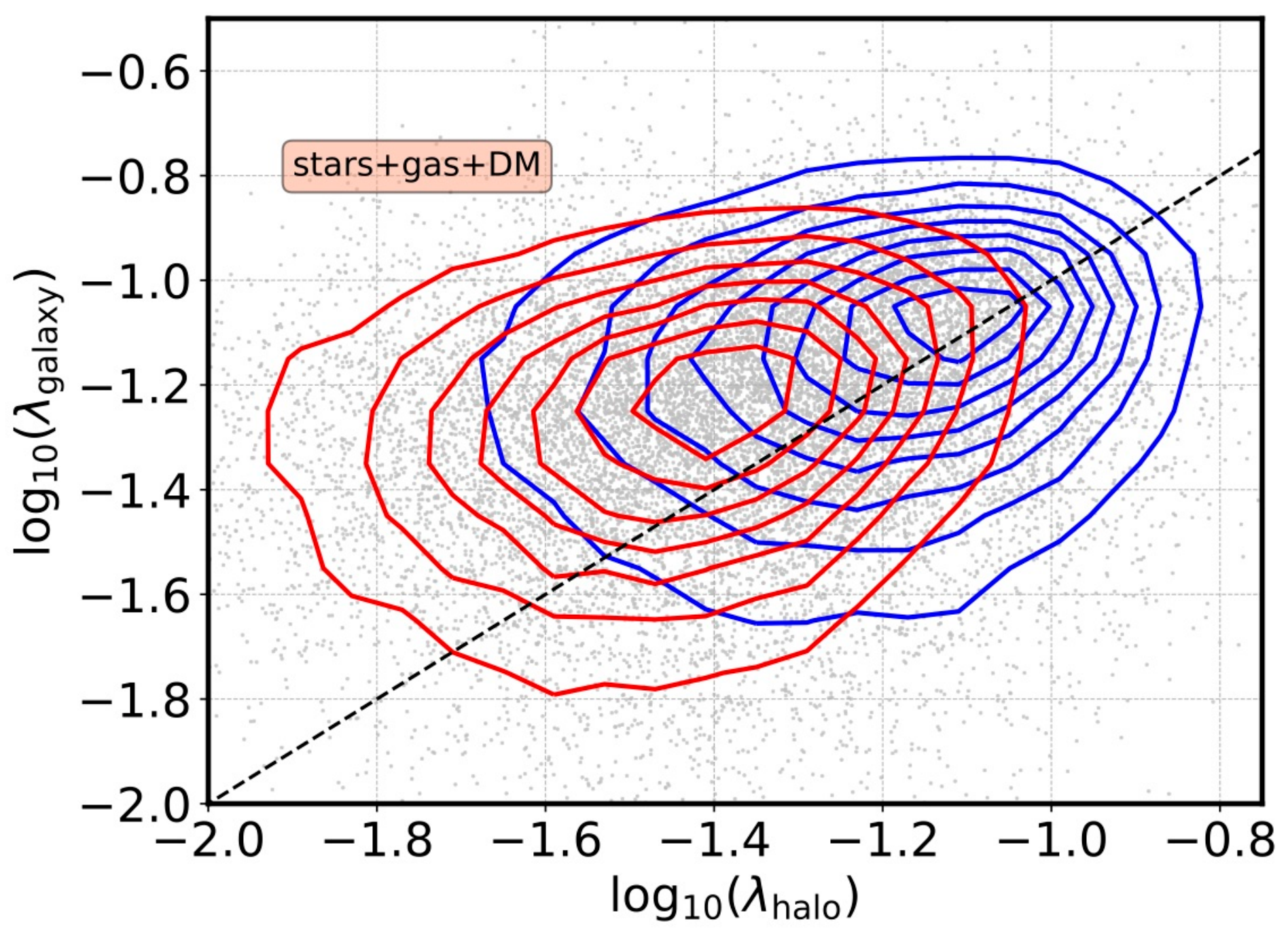}
\includegraphics[width=0.9\columnwidth]{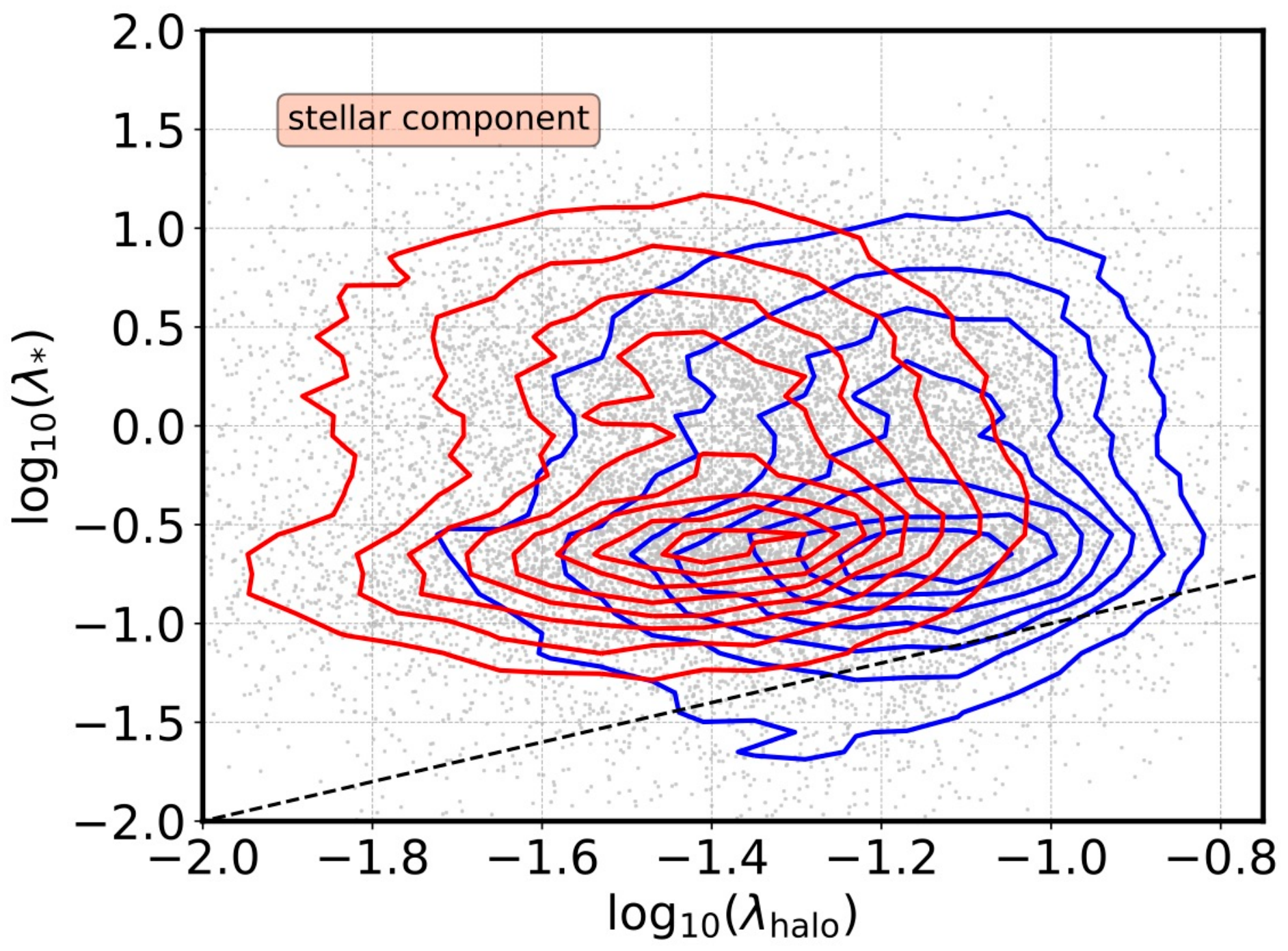}
\caption{The correlation between halo spin and galaxy spin for central galaxies. In the upper panel, the total galaxy spin, $\lambda_{\rm galaxy}$, is computed with all DM, gas and stellar-mass particles. In the lower panel, only the stellar component is considered ($\lambda_*$). In both panels, red contours show the distribution for the oldest haloes, whereas blue contours represent the youngest population (20$\%$ subsets).}
\label{fig:gsb1}
\end{figure}

\begin{figure}

\includegraphics[width=0.97\columnwidth]{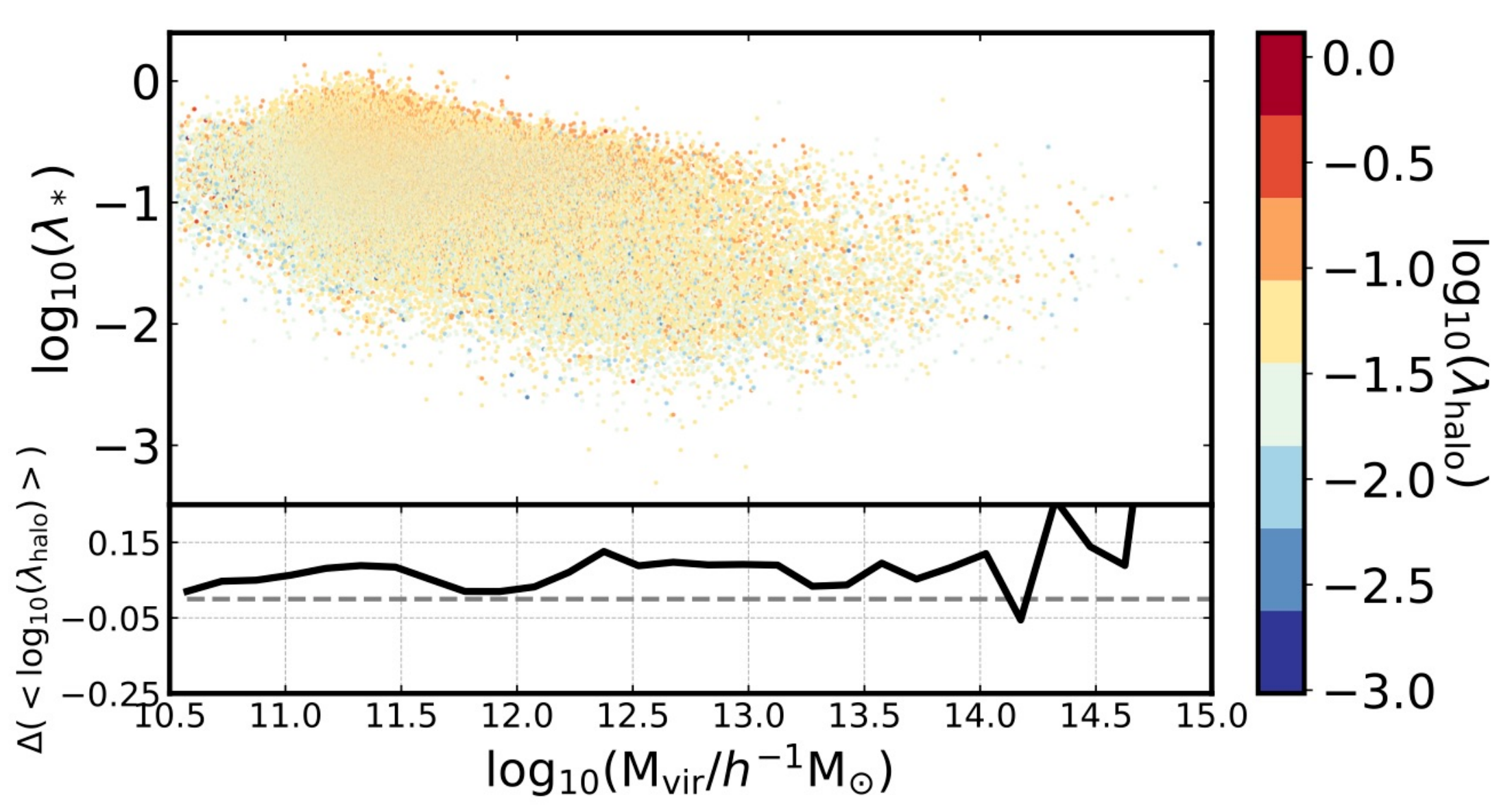}
\caption{The distribution of the stellar spin, $\lambda_*$, as a function of halo mass. The colour code shows the spin of the hosting halo, $\lambda_{\rm halo}$. In the subplot, the difference between the median $\log_{10}(\lambda_{\rm halo})$ in $50\%$ subsets of $\lambda_*$ is shown for reference, i.e., $<\log_{10}(\lambda_{\rm halo})^{(2)}> - <\log_{10}(\lambda_{\rm halo})^{(1)}>$ (where 1 and 2 represent the bottom and top subsets, respectively). }
\label{fig:spin}
\end{figure}

The dependence of galaxy clustering on secondary halo properties has mostly been probed with real data using galaxy properties that are expected to be  related to the assembly history of haloes (i.e., colour or star formation history, the so-called galaxy assembly bias effect). However, the large magnitude of the secondary dependence measured from N-body simulations on other halo properties such as spin (see the recent measurements of \citealt{SatoPolito2019} and \citealt{Johnson2019}) suggests alternative routes for the manifestation of the secondary halo bias effect on the galaxy population. In this section, we discuss the potential existence of a {\it{galaxy spin bias}} effect and its detectability with real data. 

The first aspect to investigate is the transmission of the angular momentum or spin of the halo to the central galaxy. Galaxies form when baryons collapse towards the centres of rotating haloes (and subsequently cool and condensate), so it is conceivable that their angular momenta carry some information about the angular momenta of the hosting haloes\footnote{In semi-analytic models, in fact, it is common to assume that $\lambda_{\rm galaxy} \propto \lambda_{\rm halo}$ (see, e.g., \citealt{Somerville2008, Guo2011, Benson2012}).}. This initial correlation between the angular momentum of baryons and DM is, however, threatened by several physical processes that can take place during the lifetime of galaxies, including ``wet compaction" and mergers (see, e.g., \citealt{Jiang2019, Stevens2017}).

\begin{figure}
\includegraphics[width=\columnwidth]{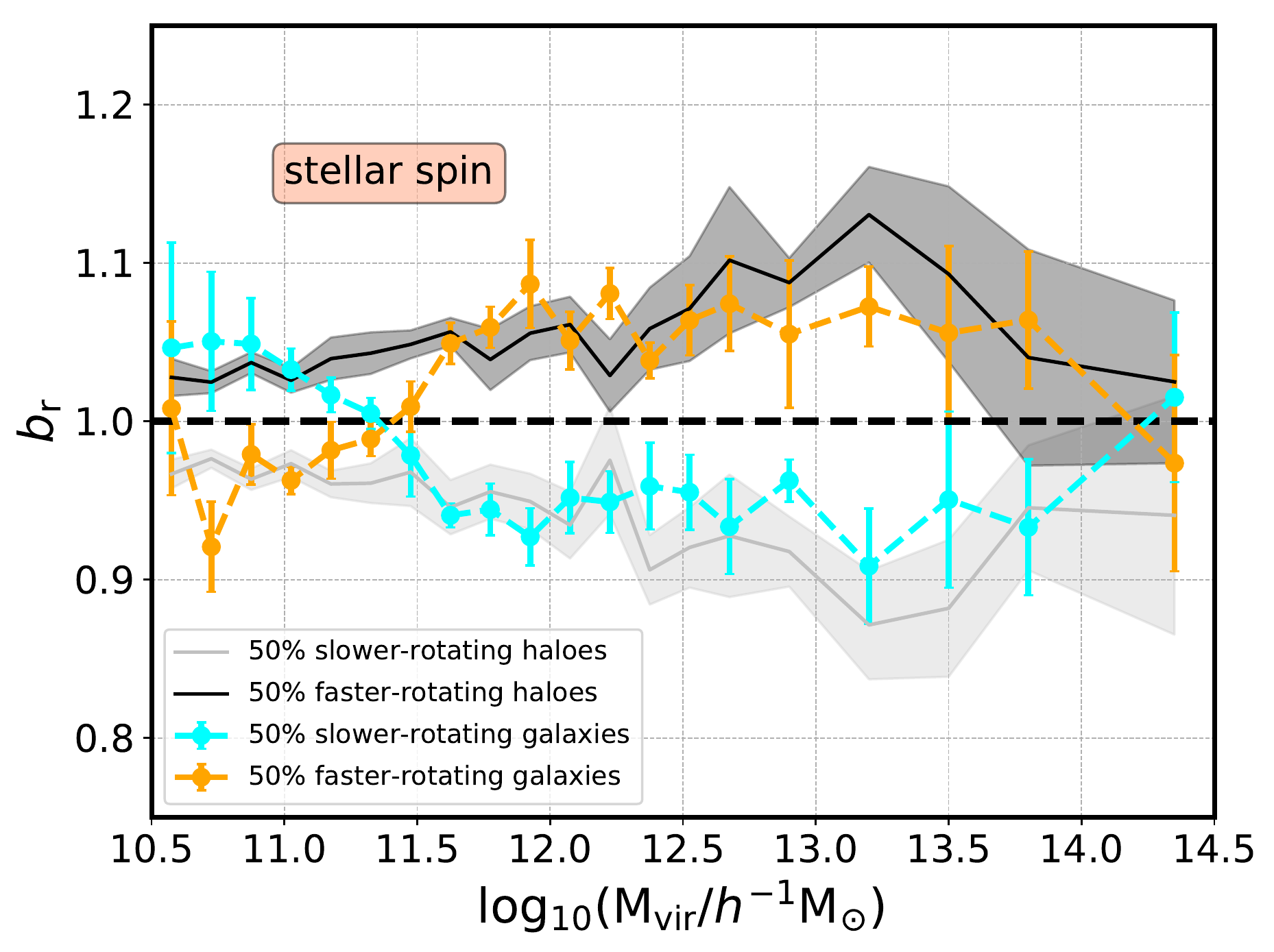}
\caption{The dependence of galaxy bias on the galaxy stellar spin ($\lambda_{*}$) as a function of halo mass in the same format of Figure~\ref{fig:gab1}. Orange symbols represent the 50$\%$ subset of galaxies with higher values of spin, whereas cyan symbols show the remaining 50$\%$ lower-spin subset. For reference, the secondary halo spin bias effect is shown in the background. The error bars represent the jackknife uncertainties obtained from a set of 8 divisions.}
\label{fig:gsb2}
\end{figure}

Figure~\ref{fig:gsb1} displays, in the upper panel, the relation between the total spin of central galaxies ($\lambda_{\rm galaxy}$, which takes into account DM, gas, and stars) and the spin of their hosting haloes, for the youngest and oldest haloes (subsets of 20$\%$ of the entire population). Figure~\ref{fig:gsb1} shows that, as expected, the halo--galaxy correlation is conserved when DM is taken into account: faster-rotating haloes clearly tend to harbour faster-rotating central galaxies, but the scatter in this correlation is significant. The separation of the hosting halo population by age indicates that: 1) younger haloes typically rotate faster, 2) the scatter in the $\lambda_{\rm galaxy}$--$\lambda_{\rm halo}$ relation is similar for older and younger haloes, and 3) the $\lambda_{\rm galaxy}$--$\lambda_{\rm halo}$ relation approximates the 1:1 relation for younger haloes. Note that the third point might be a consequence of older haloes having more time to undergo merging processes that can weaken the correlation. The same mechanism could also explain the occupancy variations of haloes measured in Illustris and IllustrisTNG. At fixed halo mass, younger haloes tend to host a larger number of satellites than older haloes, which, again, might be due to the effect of mergers \citep{Artale2018,Bose2019}.

The fact that a correlation between halo spin and galaxy total spin exists is not surprising (since this is dominated by DM). In the lower panel of Figure~\ref{fig:gsb1}, we show how this correlation appears to vanishes when only
the stellar component of the galaxy is taken into account (i.e., the stellar spin, $\lambda_{*}$). This result is in agreement with recent findings from \cite{Jiang2019}, who reported no correlation between these quantities using the NIHAO and VELA zoom-in hydro-cosmological simulations\footnote{\cite{Jiang2019} actually assume a radius of 0.1 R$_{\rm vir}$ in the computation of the galaxy stellar spin, under the assumption that this radius typically corresponds to the stellar half-mass radius. We have checked that our conclusions remain unaltered when this radius is adopted.}. These simulations, however, only contain 13 and 34 moderate-to-high-mass central galaxies across a wide redshift range, respectively. Our results thus add a strong statistical sense to their claim.

Although the correlation shown in the lower panel of Figure~\ref{fig:gsb1} is very weak, in Figure~\ref{fig:spin} we show that it is enough to produce a small separation in $\lambda_{\rm halo}$ when subsets are selected in $\lambda_{*}$. Figure~\ref{fig:spin} employs the same format as Figure~\ref{fig:distributions} to show $\lambda_{*}$ vs. M$_{\rm vir}$, but it plots $\lambda_{\rm halo}$ as a secondary halo dependency. The subplot displays a non-negligible positive $\Delta \log_{10}(\lambda_{\rm halo})$ between subsets that opens the door for a transmission of the halo spin bias effect to the galaxy population.  
In Figure~\ref{fig:gsb2}, we show the relative bias for high- and low-$\lambda_{*}$ galaxies as a function of halo mass, in the same format as Figure~\ref{fig:gab1}. In the background, we show the spin bias effect on haloes, which in IllustrisTNG300 is characterised by a progressive increase of signal as a function of halo mass. Figure~\ref{fig:gsb2} presents a clear measurement of galaxy spin bias that overall follows the halo spin bias signal.

Only at the very-low-mass end does the galaxy spin bias signal deviates from the halo spin bias signal in IllustrisTNG300. An inversion in the galaxy spin bias signal is observed at $\log_{10} ({\rm M_{vir}}/ h^{-1} {\rm M_{\odot}}) \sim 11.5$, in agreement with previous halo spin bias results \citep{SatoPolito2019, Johnson2019}. By selecting only haloes that contain a central galaxy, we are effectively removing unoccupied haloes and haloes that harbour satellite galaxies (along with haloes 
whose central galaxies fall below our resolution limit). The haloes that host satellites are identified as non-distinct haloes in N-body simulations (i.e., ``subhaloes", note the different meaning of this concept in this work), so their inclusion appears to be connected with the tensions found with respect to previous measurements of halo spin bias. Also, we have checked that by only selecting central galaxies we are effectively neglecting a small population of very-low-concentration haloes. These tensions will be fully addressed in Tucci et al. (in prep.).

\section{Discussion and conclusions}
\label{sec:discussion}

In this paper, we use the IllustrisTNG300 simulation box at redshift $z=0$ to evaluate the transmission of the secondary halo bias signal to the central galaxy population. The IllustrisTNG simulations incorporate important improvements as compared to the previous Illustris boxes, such as a new implementation of galactic winds, black-hole-driven kinetic feedback at low accretion rates, and the inclusion of magneto-hydrodynamics \citep{Weinberger2017,Pillepich2018}. Importantly, the IllustrisTNG300 box that we use here spans 205 $h^{-1}$Mpc on a side with a mass resolution below $10^8 h^{-1} {\rm M_{\odot}}$ for both baryons and DM particles, which increases considerably the statistical significance of our analysis as compared to previous measurements (e.g., \citealt{Xu2018}). The volume and resolution of IllustrisTNG300 allow us to analyse the 
clustering of central galaxies in haloes within the mass range $10.5 \lesssim \log_{10} ({\rm M_{vir}}/ h^{-1} {\rm M_{\odot}}) \lesssim 14.5$.

On the halo side, our analysis of IllustrisTNG300 yields an interesting result regarding the secondary dependence of halo bias on halo spin. Contrary to recent findings using several N-body simulations \citep{SatoPolito2019,Johnson2019}, we detect no inversion of the signal at the low-mass end (i.e., faster-rotating haloes are in IllustrisTNG300 more tightly clustered than slower rotators across the entire mass range). The fact that this feature is not present in IllustrisTNG300 might be due to the inclusion of subhaloes (i.e., haloes that live inside larger haloes in N-body numerical simulations) in the analysis or the presence of a small population of low-concentrated haloes. One possibility currently under investigation is that the ``spin crossover" is somehow a reflection of the well-established concentration crossover (even if they manifest themselves on different mass scales). Differences in the distributions of halo concentrations as a function of halo mass between different simulations might explain this tension (Tucci et al., in prep.). 

In this work, we adopt a simple definition of {\it{galaxy assembly bias}} that directly reflects the halo assembly bias effect: the dependence of the clustering of central galaxies on the formation history (or concentration) of hosting haloes at fixed halo mass. We have shown that this effect exists in various degrees when the central galaxy population is split by stellar mass, total (including DM) velocity dispersion, colour, sSFR, subhalo size, and surface density, over a wide halo mass range. We have also analysed the dependence of the signal on stellar mass at fixed M$_{\rm vir}$ (to the extent that the IllustrisTNG300 statistics permit). This implicit dependency produces different effects depending on the galaxy property used to split the galaxy population. The galaxy assembly bias signal is stronger in higher-mass galaxies for colour, sSFR, and subhalo size, and fairly independent of stellar mass for galaxy size. The surface density, on the other hand, exhibits an inversion of the galaxy assembly bias signal. 

Among the different galaxy properties considered, galaxy size emerges as a special case. At fixed halo mass, larger galaxies are more tightly clustered than smaller galaxies, but this effect, in contrast to the rest of the dependencies, seems to be uncorrelated with halo formation time.
Our results (even at ``fixed" stellar mass) seem also in tension with $z=0$ measurements from \cite{Hearin2019} indicating that small SDSS galaxies cluster much more strongly than large galaxies of the same stellar mass. It is noteworthy, however, that even though the difference observed between the two SDSS populations is strong on small scales, it seems to vanish at larger distances (i.e. $\gtrsim$ 10 $h^{-1}$Mpc). Note also that several other authors have found weak or no dependence of the mass--size relation on environment (see e.g., \citealt{HuertasCompany2013A}). Further investigation is required to understand these comparisons. Interestingly, we do find the inverse trend (smaller objects being more biased) when the subhalo size is used to split the population.  

Our results are complementary to those presented in \citet{Bose2019}, who also use lllustrisTNG300. The authors study the dependence of the galaxy content of haloes (i.e., {\it{halo occupancy variations}}) on several secondary halo properties at fixed halo mass.  Their results indicate a strong dependence of the average number of satellites on formation time, concentration and environment, at fixed halo mass. These occupancy variations of satellites imply the existence of galaxy assembly bias on the 1-halo term, whereas our study on central galaxies reflects the 2-halo term component of the effect. 

Our findings also align well with the galactic and halo conformity signal found in the Illustris simulation. \citet{Bray2016} show the existence of a strong correlation in the colours of galaxies residing in neighbouring haloes (between 3 and 10 $h^{-1}$Mpc) at either fixed stellar or halo mass (i.e., the so-called 2-halo conformity). This environmental dependence is connected to the galaxy assembly bias signal shown in our Figure~\ref{fig:gab1}, where redder galaxies are more tightly clustered than bluer galaxies at fixed halo mass, due to the different formation histories of their hosting haloes. 

We have also addressed for the first time the transmission of the halo spin bias signal to the central galaxy population, an effect that we call {\it{galaxy spin bias}}. We have shown that, 
even though central galaxy spin (the baryonic component) apparently retain little information about the total halo spin, the correlation is enough to produce a significant galaxy spin bias signal.   

\begin{figure}
\includegraphics[width=\columnwidth]{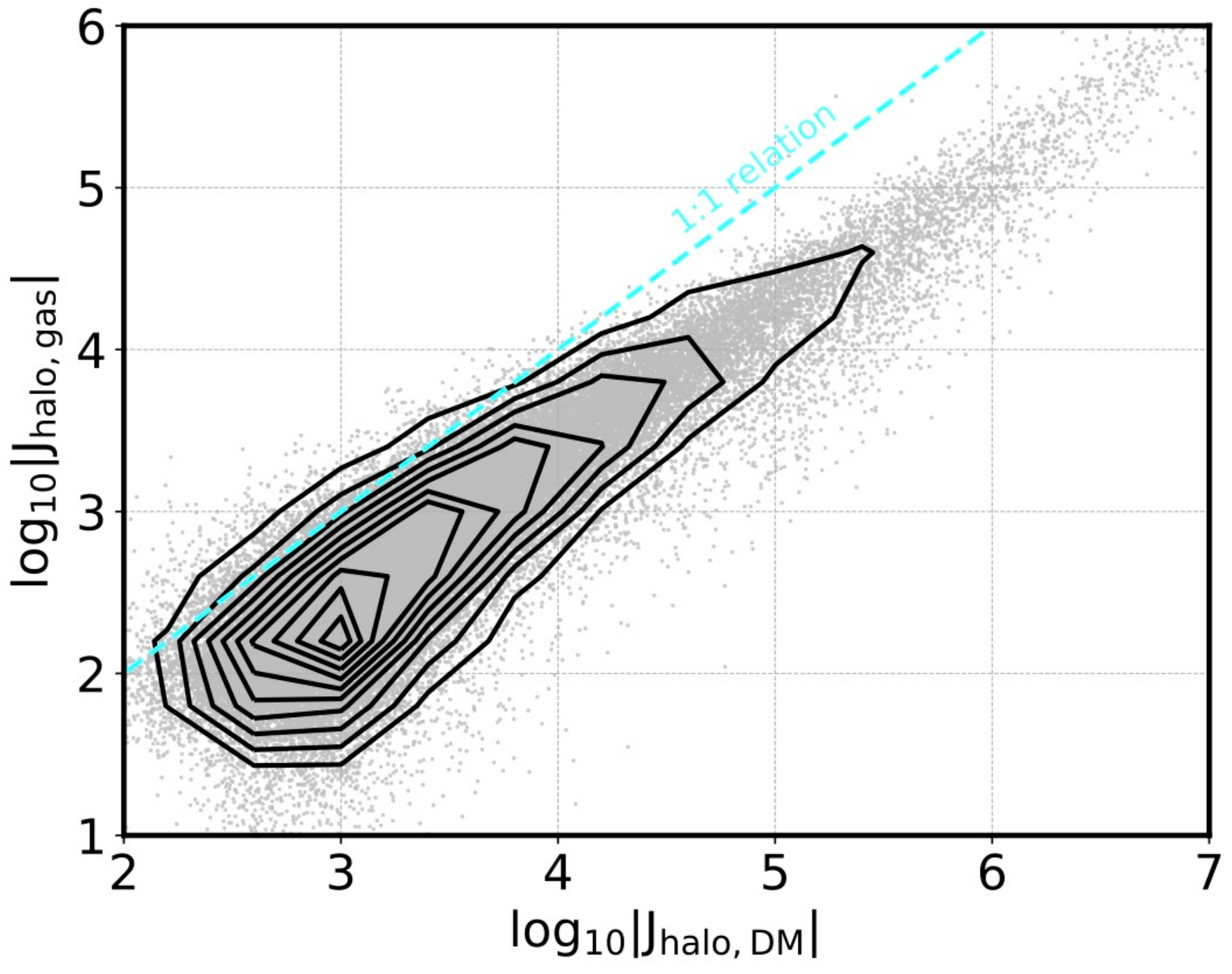}
\caption{The correlation between the gas and the DM components of the angular momenta of haloes within the virial radius R$_{\rm vir}$. Contours show the distribution of all haloes in IllustrisTNG300 with M$_{\rm vir} > 10^{10.5}$ $h^{-1}$ M$_{\odot}$, whereas dots represent a randomly selected subset containing 10$\%$ of this population. The dashed line indicates the one-to-one relation.}
\label{fig:j_halo}
\end{figure}

Although IllustrisTNG300 predicts the existence of central galaxy spin bias (something that could be directly probed by mapping the velocities of stars and gas), the idea of an observational detection of the halo spin bias signal is even more appealing. The effect of halo spin bias is particularly large at the high-mass end, i.e., for group- and cluster-size haloes. An observational detection of the halo spin bias signal would require a sizeable sample of well-identified groups/clusters with good measurements of their total masses (halo masses), along with a reliable proxy for halo spin. Ideally, one would also like the sample to expand a wide range of cluster masses for statistical reasons, even though the effect must be studied ``at fixed halo mass". In this sense, smaller groups are particularly challenging, both in terms of their identification and their mass determination through dynamical or weak-lensing methods.   

Figure~\ref{fig:j_halo} displays the tight correlation that we find between the DM angular momentum of the halo and the angular momentum of the gas that orbits inside it. This correlation suggests that the rotation of the intra-cluster gas can be used to determine halo spin. In fact, several methods have been proposed to measure this rotating intra-cluster gas. On the one hand, the neutral component of the gas can be observed through the 21-cm emission (i.e., the 21-cm hyperfine transition of neutral hydrogen), which will be accessible for millions of objects in the near future thanks to the Square Kilometer Array project (SKA\footnote{https://www.skatelescope.org})
and other similar radio surveys. Moreover, the rotation of the ionized component (i.e., the hot gas in the intra-cluster medium, ICM) can be measured through the kinetic Sunyaev-Zeldovich (kSZ) effect, i.e., the ``comptonization" of the photons of the cosmic microwave background (CMB) as they propagate through galaxy clusters due to the motion of the cluster as a whole. A purely rotational motion of the ICM would produce a detectable dipole pattern, which again can be used as a proxy for halo spin (see \citealt{Cooray2002,Chluba2002,Baldi2018}).

Upcoming surveys such as Euclid\footnote{https://www.euclid-ec.org}, the Dark Energy Spectroscopic Instrument (DESI\footnote{https://www.desi.lbl.gov}), the Javalambre Physics of the Accelerated Universe Astrophysical Survey (J-PAS\footnote{http://www.j-pas.org}), the  Prime Focus Spectroscopy survey (PFS\footnote{https://pfs.ipmu.jp}),  the Large Synoptic Spectroscopic Telescope (LSST\footnote{https://www.lsst.org}), or SPHEREx\footnote{http://spherex.caltech.edu}, to name but a few, will map the large-scale structure of the Universe (LSS) over huge cosmological volumes using a variety of galaxy populations as LSS tracers. The enhanced photometry and tracer density that these experiments will provide can improve the quality of galaxy measurements and the detection and characterisation of clusters. More importantly, their detailed weak-lensing maps can crucially reduce the uncertainties in the determination of halo masses, one of the most delicate aspects involved in probing the assembly bias effect (see, e.g., \citealt{Niemiec2018}).

\section*{Acknowledgments}

ADMD and BT thanks FAPESP for financial support. MCA acknowledges financial support from the Austrian National
Science Foundation through FWF stand-alone grant P31154-N27. LRA thanks both FAPESP and CNPq for financial support. FR has been supported by Agencia Nacional de Promoci\'on Cient\'ifica y Tecno\'oogica (PICT 2015-3098), 
the Consejo Nacional de Investigaciones Cient\'{\i}ficas y T\'ecnicas (CONICET, Argentina) and the Secretar\'{\i}a
 de Ciencia y Tecnolog\'{\i}a de la Universidad Nacional de C\'ordoba (SeCyT-UNC, Argentina).

We also thank S. Vitenti, M. Penna-Lima and C. Doux for the use of their Numerical Cosmology package, {\em{NumCosmo}}\footnote{\url{https://numcosmo.github.io/}}.

\bibliography{./paper}



\label{lastpage}

\end{document}